\documentclass[12pt,letterpaper]{article}
\usepackage{jheppub-modified}

\usepackage{graphicx}
\usepackage{bbm}
\usepackage{amsmath}
\usepackage{amssymb}
\usepackage{mathtools}
\usepackage{amsthm}
\usepackage{marvosym}
\usepackage{dsfont}
\usepackage[labelformat=simple]{subcaption}
\usepackage{enumitem}
\usepackage{multirow}

\allowdisplaybreaks

\definecolor{dark-gray}{gray}{0.20}
\definecolor{gray}{gray}{0.30}
\definecolor{light-gray}{gray}{0.80}
\definecolor{dark-red}{rgb}{0.7,0,0}
\definecolor{dark-green}{rgb}{0.1,0.4,0}
\definecolor{dark-blue}{rgb}{0.3,0.3,0.7}
\definecolor{light-blue}{rgb}{0.8,0.8,1}
\definecolor{swamp}{RGB}{240, 199, 197}

\hypersetup{
	colorlinks=true,
	linkcolor=dark-blue,
	citecolor=dark-red,
	urlcolor=dark-green,
	linktoc=page
}

\numberwithin{equation}{section}

\theoremstyle{remark}

\newtheoremstyle{named}{}{}{\itshape}{}{\bfseries}{.}{.5em}{#3}
\theoremstyle{named}

\newcommand{\be}{\begin{equation}}
\newcommand{\ee}{\end{equation}}
\def\be{\begin{equation}}
\def\ee{\end{equation}}
\def\bea{\begin{eqnarray}}
\def\eea{\end{eqnarray}}
\newcommand{\beq}{\begin{equation}}  \newcommand{\eeq}{\end{equation}}
\newcommand{\bal}{\begin{aligned}}   \newcommand{\eal}{\end{aligned}}
\def\beqa{\begin{eqnarray}}
\def\eeqa{\end{eqnarray}}

\newcommand{\brap}[1]{{\left( {#1} \right)}}
\newcommand{\bras}[1]{{\left[ {#1} \right]}}

\newcommand{\brav}[1]{{\left| {#1} \right|}}
\newcommand{\bravv}[1]{{\left\Vert {#1} \right\Vert}}

\newcommand{\rmd}{\mathrm{d}}

\newcommand{\cF}{\mathcal{F}}

\newcommand{\cK}{\mathcal{K}}

\newcommand{\cO}{\mathcal{O}}

\newcommand{\cT}{\mathcal{T}}

\newcommand{\cV}{\mathcal{V}}

\title{\centering Sharpening the Supersymmetric\\ Axion Weak Gravity Conjecture}

\author{Muldrow Etheredge$^{1}$,}
\author{Matthew Reece$^{2}$,} 
\author{Tom Rudelius$^{3}$,} 
\author{Christopher Tudball$^{3}$}
\affiliation{${}^{1}$Max-Planck-Institut f\"ur Physik (Werner-Heisenberg-Institut), Garching, 85748, GER}
\affiliation{${}^{2}$Leinweber Institute for Theoretical Physics at Harvard, Harvard University, Cambridge, MA 02138, USA}
\affiliation{${}^{3}$Department of Mathematical Sciences, Durham University, Durham, DH1 3LE, UK}

\emailAdd{muldrow.etheredge@mpp.mpg.de}
\emailAdd{mreece@g.harvard.edu}
\emailAdd{thomas.w.rudelius@durham.ac.uk}
\emailAdd{christopher.a.tudball@durham.ac.uk}

\preprint{MPP-2026-93}

\abstract{
The Axion Weak Gravity Conjecture provides one of the most effective quantum gravity tools for constraining particle physics and cosmology, but it has long been thought of as a slightly fuzzy statement: given an axion with decay constant $f$ there should exist an instanton of charge $n$ and action $S$ with $fS/|n|$ at most an order-one number in Planck units. Recent work related to axion wormholes motivated a specific order-one coefficient, $\frac{fS}{\brav{n}} \leq \frac{\pi}{2 \kappa_d} \sqrt{\frac{d-1}{d-2}}$. In this work, we verify this bound in various axion sectors across the string landscape using three complementary approaches. In the process, we derive even tighter bounds on instantons in such sectors. For example, we argue that supersymmetric instantons in 4d satisfy the stronger bound of $\frac{fS}{\brav{n}}\leq \frac 1{\kappa_4}\sqrt{\frac{7}{2}}$.}

\begin{document}
\hypersetup{pageanchor=false}
\makeatletter
\let\old@fpheader\@fpheader

\makeatother
\maketitle

\setcounter{tocdepth}{2}

\section{Introduction}
\label{sec:intro}

Among the vast landscape of quantum gravity conjectures, the axion version of the Weak Gravity Conjecture (WGC) \cite{ArkaniHamed:2006dz} offers one of the most promising routes for bridging the gap between quantum gravity theory and experiment. Most famously, the axion WGC has been shown to place meaningful bounds on simple models of natural inflation \cite{ArkaniHamed:2006dz}, including multi-field models such as $N$-flation \cite{liddle:1998jc, dimopoulos:2005ac} and axion alignment \cite{Kim:2004rp,Rudelius:2014wla, Rudelius:2015xta, Montero:2015ofa, Brown:2015iha, Heidenreich:2015wga}. The conjecture also has interesting consequences for the QCD axion~\cite{Choi:2015zra, Heidenreich:2016jrl, Heidenreich:2021yda, Reece:2023czb, Seo:2024zzs}, dark matter~\cite{Hebecker:2018ofv,Cicoli:2021gss, Sheridan:2024vtt}, quintessence~\cite{Ibe:2018ffn, Shiu:2026edl}, and early dark energy \cite{Poulin:2018dzj, Poulin:2018cxd, Rudelius:2022gyu}.

The axion WGC holds that in a consistent theory of quantum gravity in $d$ dimensions, any axion $\theta \simeq \theta + 2 \pi$ must have a charged instanton of instanton number $n$ and instanton action $S$ satisfying
\begin{equation}
\frac{f S}{\brav{n}} \leq c M_{\mathrm{Pl};d}^{\frac{d-2}{2}}\,,
\label{singleaxWGC}
\end{equation}
where $f$ is the axion decay constant, $M_{{\rm Pl;}d}$ is the reduced Planck scale in $d$ dimensions,%
\footnote{
    Throughout this paper, we set $8 \pi G_d = M_{\mathrm{Pl};d}^{2-d} = \kappa_d^2$, with the Einstein-Hilbert term normalized to $\frac{1}{2\kappa_d^2} \int d^dx\,\sqrt{-g}\mathcal{R}$. When convenient, we set $M_{\mathrm{Pl};d}=1$.
} %
and $c$ is an order-one constant.

In the past decade, the precise value of the order-one coefficient $c$ has been a source of much discussion and debate. The ordinary WGC for particles \cite{ArkaniHamed:2006dz} was originally motivated by the requirement that non-BPS black holes can decay by emitting charged particles.
Consequently, the analogous order-one coefficient is determined by the charge-to-mass ratio of an extremal black hole, and the WGC bound is simply the  extremality bound. See, e.g.,  \cite{Harlow:2022ich, Rudelius:2024mhq, Palti:2019pca} for reviews. 

For instantons charged under axions, however, there is no analogue of the Hawking evaporation process, so the original motivation for the WGC breaks down.
Furthermore, there are multiple types of semiclassical gravitational instantons, hence there are \emph{a priori} multiple options for the coefficient $c$.

The question of fixing the coefficient $c$ has been a longstanding open problem. In the recent works \cite{DiUbaldo:2026rly, Maldacena:2026jqd}, it was argued that in a theory of Einstein gravity coupled to an axion of decay constant $f$, a failure of semiclassical gravity can be cured by instantons satisfying
\begin{equation} 
\frac{f S}{\brav{n}} \leq \frac{\pi}{2} \sqrt{\frac{d-1}{d-2}} M_{\mathrm{Pl};d}^{\frac{d-2}{2}}\,.
\label{sharpenedaxWGC}
\end{equation}
While there could be other effects that also restore the validity of semiclassical gravity, this suggests the coefficient $c = \frac{\pi}{2} \sqrt{\frac{d-1}{d-2}}$ as a strong candidate for the axion WGC bound \eqref{singleaxWGC}, and we will refer to this bound henceforth as the \emph{sharpened axion WGC}. It is saturated by {\em half} the action of the Giddings-Strominger wormhole~\cite{Giddings:1987cg}, which was previously proposed to define the axion WGC in~\cite{Montero:2015ofa}. This is no coincidence: the argument of~\cite{DiUbaldo:2026rly, Maldacena:2026jqd} avoids pathologies associated with such wormhole solutions by requiring that a different instanton be unsuppressed at a specific imaginary distance in field space. 

A sharp bound of the form~\eqref{sharpenedaxWGC} requires a clear specification of the meaning of the instanton action $S$. In general, instanton effects take the form $A \exp(-S)$, so there is always an ambiguity from a rescaling of the prefactor $A \mapsto \lambda A$ accompanied by an additive shift $S \mapsto S + \log \lambda$. However, in many cases of interest, there is a controlled semiclassical expansion in which $S$ corresponds to a classical action, for example, of the Yang-Mills instanton solution or of the worldvolume of a charged object. Our interest is mostly in cases where the instanton arises from a wrapped Euclidean worldvolume, and the action is specified by the mass or tension of the charged object times the wrapped volume. In these cases, $S$ has a clear meaning, at least at leading order in an expansion in large volumes and weak couplings. When $S$ becomes small, it is more difficult to attach a meaning to the axion WGC, as the instanton expansion breaks down and certain states become light at least for some values of the axion field~\cite{Stout:2020uaf}. In that case, it may be that a more direct analysis of the imaginary distance bound of~\cite{DiUbaldo:2026rly, Maldacena:2026jqd} and the gravitational path integral can lead to a quantitative constraint without the need to refer to an instanton action. For now, we will restrict our attention to cases with controlled semiclassical expansions.

In this paper, we use three complementary approaches to verify and strengthen the bound~\eqref{sharpenedaxWGC} in various axion sectors in the quantum gravity landscape. First, we review the argument of \cite{Brown:2015iha}, which translates higher-form versions of the WGC into geometric bounds on Calabi-Yau manifolds. By compactifying different string theories on these same manifolds, these geometric bounds lead to axion WGC bounds for various axion sectors. This approach has the benefit that it applies to both BPS and non-BPS instantons; it has the drawback of relying on the assumption of the WGC for higher-form gauge fields.

Next, we develop a novel approach to strengthen the axion WGC for BPS (holomorphic) instantons in 4d supergravity theories whose K\"ahler potential takes the form
\begin{equation}
\cK = - \log(\cF(t^i))\,,
\end{equation}
for $\cF(t^i)$ a homogeneous function of the moduli $t^i$.
Such K\"ahler potentials are ubiquitous in string/M-theory compactifications, as we shall see. Furthermore, BPS  instantons are important because they are the only instantons that can contribute to the superpotential of an $\mathcal{N}=1$ theory, not merely the K\"ahler potential. From this approach, we derive the axion WGC bound:
\begin{equation} \label{eq:fSboundsqrtp}
\frac{f S}{\brav{n}} \leq \sqrt{\frac{p}{2}} M_{\rm Pl}\,,
\end{equation}
where $p$ is the degree of homogeneity of $\cF$, i.e., $\cF(\lambda t^i) = \lambda^p \cF(t^i)$. We argue that various Swampland conjectures imply that $p$ is generically a positive integer satisfying $1 \leq p \leq 7$, which ensures that the sharpened axion WGC \eqref{sharpenedaxWGC} is satisfied with room to spare. (In fact, $p = 7$ is the largest integer for which~\eqref{sharpenedaxWGC} is satisfied when~\eqref{eq:fSboundsqrtp} is saturated, so it is noteworthy that it is realized in examples in the landscape.) This second approach has the drawback that it applies specifically to BPS instantons in 4d; it has the advantage that it depends only on low-energy supergravity and is independent of the details of the UV completion.

For our third and final approach, we consider the so-called $\alpha$-vectors%
\footnote{
    The concept of these vectors originated for gradients of logarithms of masses (see, e.g.,  \cite{Palti:2017elp,Lee:2018spm, Gonzalo:2019gjp, Andriot:2020lea,   Benakli:2020pkm, Calderon-Infante:2020dhm,Etheredge:2022opl,Etheredge:2023odp, Castellano:2023jjt, Castellano:2023stg}), and has been since extended to branes or instanton actions (see, e.g.,~\cite{Lanza:2020qmt,  Etheredge:2023usk, Etheredge:2023zjk,Etheredge:2024amg, Etheredge:2024tok, Etheredge:2025ahf}). Here, we find it useful to use a positive sign-convention for instantonic $\alpha$-vectors, differing from some of the literature.
} %
of holomorphic instantons: 
\begin{equation}
\vec \alpha = \vec \nabla  \log S\,,
\label{alphavecdef}
\end{equation}
where the gradient is taken with respect to the moduli of the theory.
We show that as a simple consequence of holomorphy, these $\alpha$-vectors satisfy
\begin{equation}
\bravv{ \vec \alpha } = \frac{\brav{n}}{f S}\,.
\end{equation}
Consequently, a lower bound on the length of the $\alpha$-vector translates into an upper bound on $fS/\brav{n}$. We then derive lower bounds on lengths of instanton $\alpha$-vectors in various axion sectors within the string/M-theory landscape. This approach has the drawback that it applies only to holomorphic instantons; it has the advantage that it applies to string/M-theory compactifications in any number of dimensions.

We will see that these three approaches are perfectly complementary in that the coefficients $c$ derived by one approach perfectly match those derived by the other approaches whenever two or more approaches are simultaneously applicable. This suggests that the results of each approach may extend beyond the apparent domain of validity: for instance, bounds on BPS instantons in certain theories may extend to non-BPS instantons in those theories as well. In addition, all three of these approaches extend straightforwardly to theories with multiple axions. Where there is overlap, our results are also consistent with past explicit tabulations of $fS$ for axions in string theory, e.g., in~\cite{Cicoli:2021gss}.

While instantons are electrically charged under axion gauge fields, in 4d axion strings are magnetically charged under axion gauge fields.
Axion strings have their own version of the weak gravity conjecture, namely that in a consistent theory of quantum gravity in $d$ dimensions, any axion $\theta$ must have an axion string magnetically charged under $\theta$ satisfying
\begin{equation}\label{eq:axion-string-wgc}
    \frac{\cT}{\tilde g} \leq c'M_{\mathrm{Pl};4}
    \,,
\end{equation}
where $\cT$ is the axion string's tension, $\tilde g$ is the magnetic coupling of the axion string to the axion, and $c'$ is an order-one constant.
Our second and third approaches to strengthening the axion WGC can be used to find a value of $c'$ such that all BPS axion strings charged under the axions considered in those approaches satisfy \eqref{eq:axion-string-wgc} for said value of $c'$; we investigate this in \S\ref{ss.boundaxionstrings} and \S\ref{ss.axionstring} respectively.

The remainder of this paper is structured as follows. In \S\ref{s.Background}, we introduce our conventions and review the extension of the axion WGC to theories with multiple axion fields. In \S\ref{s.DUALITIES}-\ref{s.ALPHA}, we present each of our three approaches in turn. In \S\ref{s.DUALITIES}, we review the argument of \cite{Brown:2015iha} for the axion WGC using the duality web. In \S\ref{s.HOMOGENEOUS}, we derive bounds on BPS instantons in theories with homogeneous K\"ahler potentials. In \S\ref{s.ALPHA}, we use the connection with $\alpha$-vectors to derive bounds on BPS instantons. In \S\ref{s.INFLATION}, we comment briefly on applications of our bounds to natural axion inflation. In \S\ref{s.DISC}, we conclude with a summary and potential next steps.

\section{Background and Conventions}
\label{s.Background}

In this section, we present relevant background on the Weak Gravity Conjecture (WGC) and axions, and we introduce notation that will be used in the remainder of the paper.

\subsection{The Weak Gravity Conjecture}

Consider a theory of real massless scalar fields $\phi^a$, $a=1,...,n_s$ and $p$-form gauge fields $A_p^i$, $i=1,...,n$ coupled to gravity in $d$ spacetime dimensions:
\begin{equation}
S = \int \sqrt{-g}\, d^dx \left(\frac{1}{2 \kappa_d^2} \mathcal{R} - \frac{1}{2} g_{ab}(\phi) \partial_\mu \phi^a  \partial^\mu \phi^b - \frac{1}{2} \tau_{ij}(\phi) F_{p+1}^i \cdot F_{p+1}^j \right)\,.
\label{genact}
\end{equation}
Here, $F_{p+1} \cdot G_{p+1} = \frac{1}{(p+1)!} F_{\mu_1...\mu_{p+1}}G^{\mu_1...\mu_{p+1}}$, $g_{ab}$ is the metric on moduli space, and $\tau_{ij}(\phi)$ is the gauge kinetic matrix, which in general depends on the moduli $\phi^a$.

We work in a lattice basis, so that a $(p-1)$-brane carries integral charge $q_i \in \mathbb{Z}^n$ under the respective gauge fields $A^i$. The normalized charge is then given by contracting with the inverse of the gauge kinetic matrix:
\begin{equation}
\bravv{q}^2 = q_i \tau^{ij} q_j\,.
\end{equation}
It is useful to define the charge-to-tension vector $z_i = q_i/(T \kappa_d)$, where $T$ is the tension of the brane. The norm of such a $z$-vector is also determined by contracting with the inverse gauge kinetic matrix, $\bravv{z} = z_i \tau^{ij} z_j$

\begin{figure}
    \centering
    \includegraphics[width=0.6\linewidth]{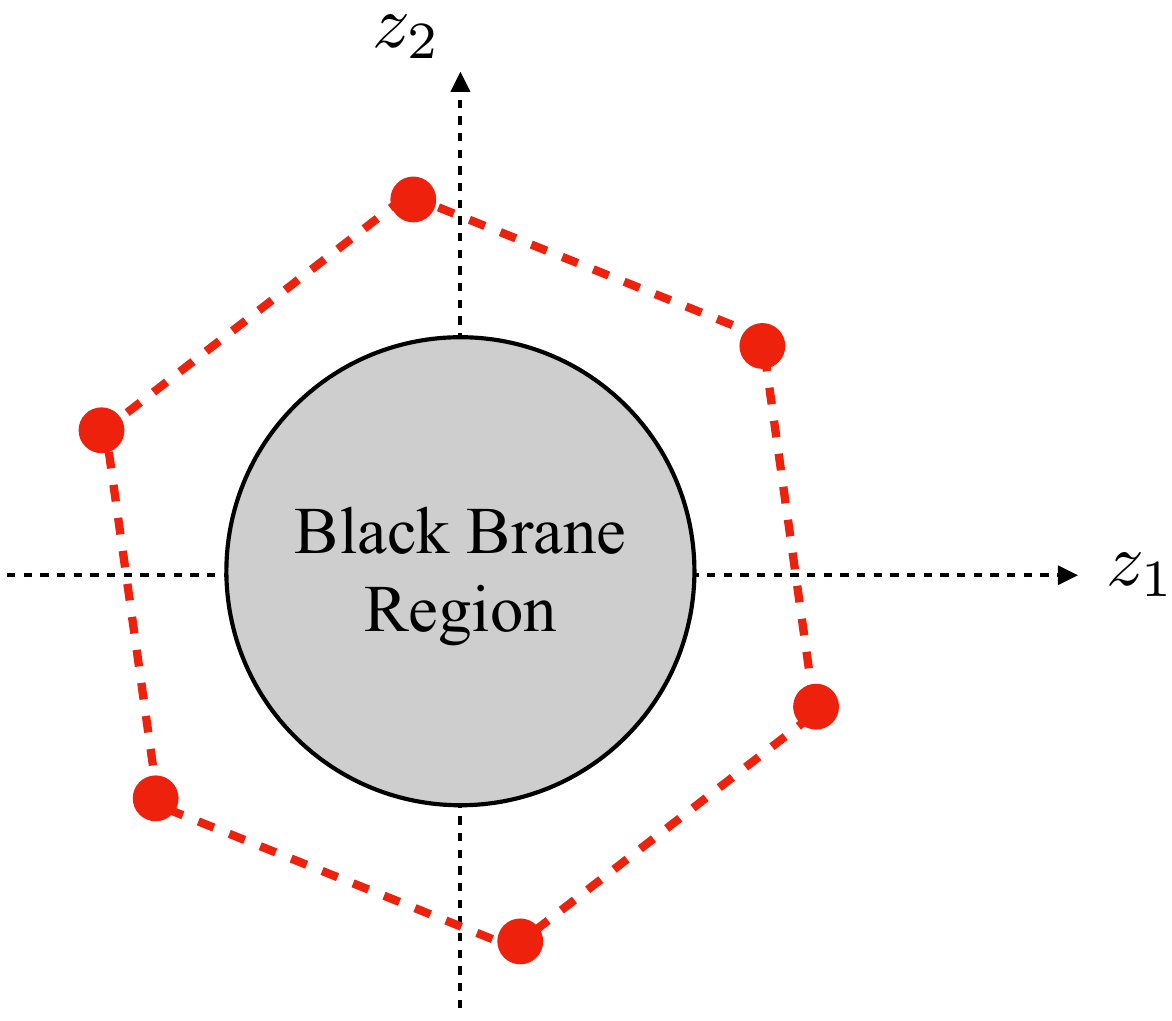}
    \caption{The convex hull condition. In theories with multiple gauge fields, the WGC is equivalent to the statement that the convex hull of the charge-to-tension vectors $z_i$ of the charged branes (red) must contain the black brane region (shaded gray).}
    \label{fig.convexhull}
\end{figure}

Given such a theory, the WGC requires a superextremal $(p-1)$-brane state in every direction of the charge lattice. As shown in \cite{Cheung:2014vva}, this can be usefully formulated as the statement that the convex hull of the charge-to-tension vectors $z_i = q_i/T$ contains the black brane region, as shown in Figure \ref{fig.convexhull}.

In general, the size and shape of the black brane region depend on the moduli $\phi^a$ and their coupling to the gauge fields through $\tau(\phi)$. The boundary of the region consists of extremal black branes, which have vanishing self-force \cite{Heidenreich:2020upe}, i.e., the long-range force between a pair of identical extremal black branes vanishes. This vanishing self-force involves a balance between the repulsive gauge force and the attractive gravitational and scalar forces. Therefore, the presence of moduli $\phi^a$ can only increase the size of the black brane region in charge-to-tension space \cite{Heidenreich:2019zkl}. This means that the black brane region will always contain the 
ball of radius
\begin{equation}
\bravv{Z}_{\rm RN} = \sqrt{\frac{p(d-p-2)}{d-2}}  \,,
\label{ZRN}
\end{equation}
which is the size of the ``Reissner-Nordstr\"om'' black brane region in a theory without massless scalar fields. The WGC thus implies that the convex hull of the charge-to-tension vectors contains this ball as well.

In a theory with a single $U(1)$, the WGC reduces to the simple expression
\begin{equation}
\frac{\brav{n}e_{p;d}}{T} \geq \bravv{Z}_{\rm ext} \geq \sqrt{\frac{p(d-p-2)}{d-2}}  \,,
\end{equation}
where $e_{p;d} = 1/\sqrt{\tau}$ is the gauge coupling of the $p$-form gauge field, $q = n$ is the charge of the brane, and $\bravv{Z}_{\rm ext}$ is the charge-to-tension ratio of an extremal black brane.

\subsection{The axion WGC}\label{ss.axionWGC}

Periodic scalar fields, also known as axions, can be usefully described as 0-form gauge fields, which couple electrically to objects of worldvolume dimension 0, also known as instantons. The axion WGC thus describes the extension of the WGC described in the previous subsection to the case $p=0$. 

For $p=0$, the action \eqref{genact} takes the form
\begin{equation}
S = \int \sqrt{-g}\, d^dx \left(\frac{1}{2 \kappa_d^2} \mathcal{R} - \frac{1}{2} g_{ab}(\phi) \partial_\mu \phi^a  \partial^\mu \phi^b - \frac{1}{2} f_{ij}(\phi) \partial_\mu  \theta^i \partial^\mu \theta^j \right) \,.
\end{equation}
Here, each $\theta^i$ has periodicity $2\pi$, i.e., $\theta^i \simeq \theta^i + 2 \pi$, and we have denoted the axion kinetic matrix by $f_{ij}$. For the case of a single axion, $f_{ij} = f^2$ is the square of the axion decay constant $f$.

The charge of an instanton under the axions $\theta^i$ is labeled by a vector $q_i \in \mathbb{Z}^n$. Its norm is given by contraction with the inverse axion kinetic matrix:
\begin{equation}
\bravv{q}^2 = q_i f^{ij} q_j \equiv \frac{1}{f_q^2}\,.
\end{equation}

By the analogy with the higher-form case $p \geq 1$, we define charge-to-action vectors
\begin{equation}
z_i = \frac{q_i}{S} \,,~~~~\bravv{z} = z_i f^{ij} z_j = \frac{1}{ f_q^2 S^2} \,.
\end{equation}
For a theory with a single axion, we have $f_{ij} =f^2$, $q=n$, and therefore
\begin{equation}
\bravv{z} = \frac{\brav{n}}{fS}\,.
\end{equation}

We would then like to define the axion WGC as the statement that the charge-to-action vectors $z_i$ of the instantons of the theory must contain the gravitational instanton region. Here, a problem arises: there are three types of gravitational instantons (extremal instantons, cored instantons, and wormholes)~\cite{Gutperle:2002km, Bergshoeff:2004fq,Bergshoeff:2004pg}, and it is not \emph{a priori} clear which of them is the correct one to use~\cite{Heidenreich:2015nta, Hebecker:2016dsw, Hebecker:2018ofv}. Naively setting $p=0$ in \eqref{ZRN} yields a trivial bound, indicating that the axion WGC is more subtle to define than its higher-form cousins. For this reason, studies of the axion WGC in string theory (e.g.,~\cite{Marchesano:2019ifh,Grimm:2019wtx}) have often focused on understanding how the theory avoids {\em parametric} violations of the bound.

Recently, the papers \cite{DiUbaldo:2026rly, Maldacena:2026jqd} argued that in a theory of an axion of decay constant $f$ coupled to quantum gravity, there should exist an instanton of charge $n$ and action $S$ satisfying 
\begin{equation}
\bravv{z} \equiv \frac{\brav{n}}{f S} \geq \bravv{Z}_{\rm GS} \equiv \frac{2}{\pi} \sqrt{\frac{d-2}{d-1}} \,.
\label{zaxboundsharp}
\end{equation}
Here, the minimal value $\bravv{Z}_{\rm GS}$ is determined by half of the charge-to-action ratio of the Giddings-Strominger wormhole~\cite{Giddings:1987cg}, as originally proposed for the axion WGC in~\cite{Montero:2015ofa}.
In theories with multiple axions, it generalizes to the statement that the convex hull of the charge-to-action vectors $z_i$ must contain the ball of radius $\bravv{Z}_{\rm GS}$ \cite{Maldacena:2026jqd}. This bound was derived as a way to avoid a pathology associated with a specific finite {\em imaginary} distance in axion field space.%
\footnote{
    This is distinct from previous work on upper bounds for finite {\em real} values of the axion in a holographic context~\cite{Hamada:2019fmc,Hamada:2020phg}, or from familiar phase transitions for finite real values of the theta parameter~\cite{Dashen:1970et, Witten:1980sp, Gaiotto:2017yup}.
} %
This argument seems to fall somewhat short of a full proof of~\eqref{zaxboundsharp}, as it leaves open the possibility that some physics other than instantons resolves the pathology associated with imaginary distances.

The result~\eqref{zaxboundsharp} was derived in the case without massless moduli. However, it is widely believed that massless moduli arise only in theories with preserved supersymmetry. By weakly breaking supersymmetry, we expect that one can give small masses to the scalar fields without substantially altering the other parameters of the theory, in which case the bound \eqref{zaxboundsharp} applies to theories both with and without massless moduli.

Nonetheless, in theories of $p$-form gauge fields with $p \geq 1$, the WGC leads to a stronger bound in theories with massless moduli than it does in theories without massless moduli~\cite{Heidenreich:2015nta}. In this paper, we will show that the same is true for the axion WGC: large classes of instantons in the string landscape and supergravity satisfy a strictly stronger bound than \eqref{zaxboundsharp}. To this end, we now present relevant details of supergravity and BPS instantons in four dimensions.

\subsection{Supergravity and BPS instantons}\label{ss.Supergravity}

In a supersymmetric theory of gravity in $d=4$ dimensions, the  scalar fields $\phi^a = t^i$ (also known as ``saxions'') and the axions $\theta^i$ pair up into complex scalar fields:
\begin{equation}
T^i = t^i - i\frac{\theta^i}{2 \pi}\,.
\end{equation}
There is thus an equal number of saxions and axions, $n=n_s$, and the saxion kinetic matrix $g_{ij}$ is related to the axion kinetic matrix $f_{ij}$ via
\begin{equation}
g_{ij} = (2 \pi)^2 f_{ij} = \frac{1}{2} \partial_i \partial_j \cK(t^i)\,,
\end{equation}
where $\cK(t^i)$ is the K\"ahler potential and $\partial_i$ denotes a derivative with respect to the modulus $t^i$. The shift symmetry of the axions ensures that (within a controlled regime of moduli space), $\cK$ depends on $\theta^i$ only through non-perturbatively small corrections.

A BPS instanton---also known as a holomorphic instanton in the supersymmetric context---is an instanton whose action is related to its charge $q_i$ via
\begin{equation}
S_q = 2 \pi q_i t^i\,.
\end{equation}
Such instantons are important, among other reasons, because only these instantons can contribute to the superpotential $W(T^i)$, which is a holomorphic function of the complex modulus $T^i$. Such contributions take the form
\begin{equation}
W(T^i) = \mathcal{A} e^{- 2 \pi q_i T^i} = \mathcal{A} e^{-S_q + i q_i \theta^i} \,,  
\end{equation}
which notably respects the axion shift symmetry.

\begin{figure}
    \centering
\includegraphics[width=0.7\linewidth]{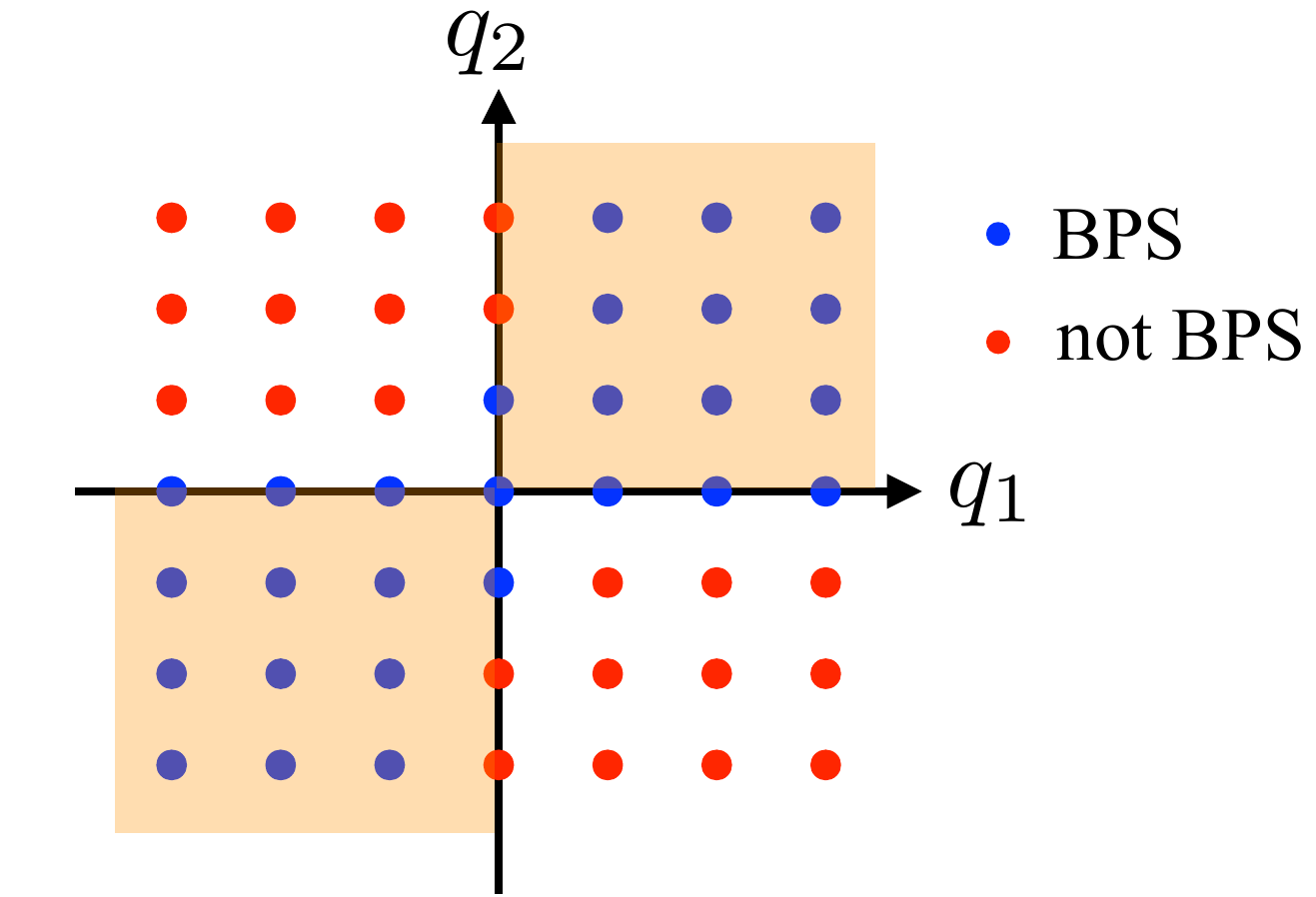}
    \caption{The BPS Cone. BPS instantons charges are constrained to lie inside the BPS cone, which may be identified geometrically with the effective cone of the compactification manifold.}
    \label{fig.BPScone}
\end{figure}

In string theory, axions often appear as holonomies of $p$-form gauge fields over $p$-cycles of compactification manifolds~\cite{Witten:1984dg,Svrcek:2006yi,Conlon:2006tq}:
\begin{equation}
\theta^i = \int_{\Sigma_p^i} C_p\,.
\end{equation}
See \cite{Reece:2025thc} for a detailed discussion of these extra-dimensional axions. In such compactifications, BPS instantons arise from charged, BPS $(p-1)$-branes wrapped over calibrated $p$-cycles. The charges of BPS instantons reside within a particular cone inside the charge lattice known (in different contexts) as the effective cone, the Mori cone, or simply the BPS cone. See Figure \ref{fig.BPScone}.

\section{Axion WGC from Dualities}\label{s.DUALITIES}

In this section, we review the argument of \cite{Brown:2015iha}, which derived bounds on various axion sectors in string compactifications using higher-form versions of the WGC and the duality web. This approach of using well-understood versions of the WGC to constrain geometry, which in turn can imply less well-understood versions of the WGC, was dubbed the ``geometric weak gravity conjecture'' in~\cite{Hebecker:2015zss}.

We begin by considering M-theory on a Calabi-Yau threefold. This theory has electrically charged particles associated with M2-branes wrapped on 2-cycles and magnetically charged strings associated with M5-branes wrapped on 4-cycles.

\begin{table}[ht]
	\centering
	\renewcommand*{\arraystretch}{1.5}
	\begin{tabular}{|ccc|ccc|c|}
		\hline
        \multicolumn{3}{|c|}{Theory A} & \multicolumn{3}{|c|}{Theory B} & 
        \\
		Theory & Branes & WGC for & Theory & Branes & WGC for & $\brav{z}_{\min}$
		\\\hline
		M-theory & M2 & 5d particles & Type IIB & F1 & 4d inst. & $\sqrt{\frac23}$
		\\
		M-theory & M2 & 5d particles & Type IIB & D1 & 4d inst. & $\sqrt{\frac23}$
		\\
		M-theory & M5 & 5d strings & Type IIB & D3 & 4d inst. & $\sqrt{\frac23}$
		\\
		Type IIB & D3 & 4d particles & Type IIA & D2 & 4d inst. & $\sqrt{\frac12}$
		\\\hline
	\end{tabular}
	\caption{Axion WGC via the duality web. After compactifying theory A on a Calabi-Yau manifold, the WGC for particles/strings yields a geometric bound on the sizes of various cycles. Compactifying theory B on this same manifold, this geometric bound translates into a WGC bound on instantons charged under axions; namely, the convex hull of the $z$-vectors of the charged instantons should contain the ball of radius $\bravv{z}_{\rm min}$.}
	\label{tab:dual}
\end{table}

In general, the WGC bound for charged $(p-1)$-branes in $d$ spacetime dimensions is moduli-dependent. However, since massless scalar fields mediate attractive long-range forces, moduli always increase the charge-to-tension ratio $Z_{\rm BH}$ of an extremal black brane relative to the charge-to-tension ratio $Z_{\rm RN}$ of an extremal Reissner-Nordstr\"om black brane. Therefore, moduli always strengthen the WGC bound, so any superextremal brane must satisfy%
\footnote{
    This differs from the value reported in \cite{Brown:2015iha} by a factor of $\sqrt{2}$; this discrepancy seems to be due to a difference in conventions in the definition of $z$.
}
\begin{equation}
\bravv{z}^2 \equiv \frac{q^2}{T^2} \geq Z_{\rm BB}^2 \geq Z_{\rm RN}^2 = \frac{p(d-p-2)}{d-2}\,.
\label{genz}
\end{equation} 
For both particles and strings in $5$ dimensions, this yields
\begin{equation}
\bravv{z} \geq \sqrt{\frac{2}{3}}\,.
\label{zboundM}
\end{equation}
In an M-theory compactification on a Calabi-Yau threefold, the WGC for particles translates into a bound on the volumes of 2-cycles, while the WGC for strings translates into a bound on the volumes of 4-cycles. Compactifying Type IIB string theory on this same Calabi-Yau, this geometric condition on 2-cycle volumes translates into bounds on worldsheet instantons and D1-brane instantons, and the bound on 4-cycle volumes translates into a bound on D3-brane instantons. In particular, we find that each of these types of instantons must satisfy
\begin{equation}
\bravv{z} = \frac{1}{f_q S_q} \geq \sqrt{\frac{2}{3}} \,,
\end{equation}
where the factor of $\sqrt{2/3}$ descends directly from the factor on the right-hand side of \eqref{zboundM}.
Note that this argument does not rely on a BPS condition: it applies not only to the holomorphic instantons but to the non-holomorphic ones as well.

A similar argument applies to D2-brane instantons charged under $C_3$ axions in Type IIA string compactification on Calabi-Yau manifolds. Compactifying Type IIB string theory on the same manifold, one obtains particles from D3-branes wrapping 3-cycles. By \eqref{genz}, these particles must satisfy
\begin{equation}
\bravv{z} \geq \frac{1}{\sqrt{2}}\,,
\end{equation}
which places constraints on volumes of 3-cycles.
Translating these constraints into Type IIA language, we find that D2-brane instantons must satisfy
\begin{equation}
\bravv{z} = \frac{1}{f_q S_q} \geq \frac{1}{\sqrt{2}}\,.
\end{equation}
These results are shown in Table \ref{tab:dual}.

\section{Homogeneous K\"ahler Potentials}\label{s.HOMOGENEOUS}

Let us suppose that the K\"ahler potential takes the general form
\begin{equation}
\cK = -  \log(\cF(t^i))\,,
\label{Khom}
\end{equation}
where $\cF(t^i)$ is a homogeneous function of degree $p$, i.e., $\cF(\lambda t^i) = \lambda^p \cF(t^i)$.

The K\"ahler metric then takes the form
\begin{equation}
\kappa_{ij} = \partial_i \partial_j \cK =  \left(\frac{\cF_i \cF_j}{\cF^2} - \frac{\cF_{ij}}{\cF} \right) \,.
\end{equation}
Here, $\cF_i = \partial_i \cF$ and $\cF_{ij} = \partial_i \partial_j \cF$. Using the homogeneity of $\cF$, we have
\begin{equation}
t^i \cF_i =  p \cF\,,~~~~ t^i \cF_{ij} = (p-1) \cF_j\,,~~~ \cF^{ij} \cF_j  = \frac{1}{p-1} t^i\,,
\label{genFid}
\end{equation}
where $\cF^{ij}$ is the inverse of $\cF_{ij}$, i.e., $\cF^{ij} \cF_{jk} = \delta^i_k$.

The inverse metric $\kappa^{ij}$ is then given by
\begin{equation}
\kappa^{ij} =  \left(\frac{1}{p-1} t^i t^j - \cF \cF^{ij} \right)\,.
\label{kappainv}
\end{equation}

\subsection{Positivity lemma}\label{ss.poslemma}

By assumption, the metric $\kappa_{ij}$ is positive-definite. Under this assumption, we claim that the modified metric
\begin{equation}
h_{ij} =  \left(\frac{p-1}{p} \frac{\cF_i \cF_j}{\cF^2} - \frac{\cF_{ij}}{\cF} \right)
\end{equation}
is positive-semidefinite.

To see this, we first note that we may write
\begin{equation}
\kappa_{ij} = h_{ij} + \xi_{ij}\,,~~~~
\xi_{ij} = \frac{1}{p} \frac{\cF_{i} \cF_{j}}{\cF^2}\,.
\end{equation}
We also note that
\begin{equation}
 t^i h_{ij}  = 0\,,
\label{genvanish}
\end{equation}
using the identities in \eqref{genFid}.

From here, the proof proceeds by contradiction. Let us suppose that $h_{ij}$ is \emph{not} positive-semidefinite. Then, there exists $v^i$ such that $v^i h_{ij} v^j < 0$. Without loss of generality, we may decompose $v^i$ as
\begin{equation}
v^i = \alpha t^i + u^i\,,~~~~\text{where } \alpha = \frac{ v^i \cF_i}{p \cF} \,.
\end{equation}
With this decomposition, we have
\begin{equation}
u^i \cF_i = (v^i - \alpha t^i )\cF_i = 0\,.
\end{equation}
We then have
\begin{equation}
0 > v^i h_{ij} v^j = (\alpha t^i + u^i) h_{ij} (\alpha t^j + u^j) = u^i h_{ij} u^j\,,
\end{equation}
where in the last equality we have used \eqref{genvanish}. From here, we may use the definition of $\xi_{ij}$ and the fact that $u^i \cF_i = 0$ to set
\begin{equation}
0 > u^i h_{ij} u^j = u^i (\kappa_{ij} - \xi_{ij}) u^j = u^i \kappa_{ij} u^j\,.
\end{equation}
This contradicts our assumption that $\kappa_{ij}$ was positive definite. We conclude that $h_{ij}$ is positive-semidefinite.

\subsection{A sharpened WGC}\label{ss.sharpenedhom}

As discussed in \S\ref{ss.Supergravity}, the metric on axion moduli space is given by
\begin{equation}
\label{eq:homogeneous-kahler-decay-constant-matrix}
f_{ij} = \frac{1}{2 (2 \pi)^2} \kappa_{ij}\,.
\end{equation}
A BPS instanton of charge $q_i$ has an action given by
\begin{equation}
S_q = 2 \pi q_i t^i\,.
\end{equation}
The associated charge/decay constant is given by
\begin{equation}
\bravv{q}^2 \equiv f_q^{-2} = q_i f^{ij} q_j\,,
\end{equation}
where $f^{ij}$ is the inverse of $f_{ij}$.

With this, we claim that the axion WGC takes the form
\begin{equation}\label{eq:homogeneous-kahler-awgc-bound}
f_q S_q \leq \sqrt{\frac{p}{2}}\,.
\end{equation}
To prove this, we first define
\begin{equation}
\tilde q^i = f^{ij} q_j\,.
\end{equation}
With this,
\begin{equation}
S_q = 2 \pi \tilde q^i f_{ij} t^j = \frac{1}{4 \pi} \frac{\tilde q^i \cF_i}{\cF}\,,
\label{Stact}
\end{equation}
while
\begin{align}
\bravv{q}^2 = \tilde q^i f_{ij} \tilde q^j = \frac{1}{2(2 \pi)^2} \left( \frac{(\tilde q^i \cF_i)^2}{\cF^2} - \frac{\cF_{ij} \tilde q^i \tilde q^j}{\cF} \right) \,.
\end{align}
We may further rewrite this as
\begin{equation}
\bravv{q}^2 = \frac{1}{2 p (2 \pi)^2} \frac{(\tilde q^i \cF_i)^2}{\cF^2} + \frac{1}{2 (2 \pi)^2} \tilde q^i h_{ij} \tilde q^j \,,
\end{equation}
where $h_{ij}$ is the matrix proven to be positive-semidefinite in \S\ref{ss.poslemma}. Using this positivity and substituting \eqref{Stact}, we have
\begin{equation}
\bravv{q}^2 \geq \frac{1}{2 p (2 \pi)^2} \frac{(\tilde q^i \cF_i)^2}{\cF^2} = \frac{ 2 }{ p} S_q^2\,.
\end{equation}
Setting $f_q = 1/\bravv{q}$, we arrive finally at
\begin{equation}
f_q S_q \leq \sqrt{\frac{p}{2}}\,,
\label{fScpbound}
\end{equation}
which completes the proof.

Note that this bound is saturated for an instanton of charge $q_i \propto \cF_i$.%
\footnote{
    Note that this $q_i$ is a moduli-dependent quantity. However, if we evaluate this charge $q_i$ at a fixed point $t_\ast$ in moduli space, making it explicitly moduli-independent, we find that it saturates the bound at the point $t=t_\ast$ in moduli space.
    It is also not, in general, a rational direction in charge space, but properly quantized instantons can come arbitrarily close to it.
} %
To see this, set $q_i = a \cF_i$. Then, using the identities \eqref{genFid} and \eqref{kappainv}, we have
\begin{equation}
\bravv{q}^2= 2\cdot (2 \pi a)^2 \cF_i \kappa^{ij} \cF_j = 2p \cdot  (2 \pi a \cF)^2 \,,~~~~S_q = 2 \pi q_i t^i = 2 \pi a p \cF \,.
\end{equation}
Together, these give
\begin{equation}
f_q S_q = \frac{S_q}{\bravv{q}} = \sqrt{\frac{p}{2}}\,,
\end{equation}
so \eqref{fScpbound} is saturated for any BPS instanton of charge $q_i \propto \cF_i$, as claimed. 

\subsection{Generalization to multiple terms}\label{ss.multiple}

Suppose now that the K\"ahler potential can be written as a sum of terms of the form in \eqref{Khom}:
\begin{equation}
\cK = - \sum_{\alpha=1}^N \log(\cF_{\alpha}(t_\alpha^i))\,,
\label{multiterm}
\end{equation}
where $\cF_{\alpha}(\lambda t_\alpha^i) = \lambda^{p_\alpha} \cF_{\alpha}( t_\alpha^i)$ is a homogeneous function of the moduli $t_\alpha^i$ of degree $p_{\alpha}$. In this case, the metric assumes a block-diagonal form, and each of the $N$ terms in the prepotential corresponds to a distinct decoupled sector. For an instanton carrying charge $q$ solely in the $\alpha$th sector, we may disregard all other terms in the sum, yielding
\begin{equation}
(f_{q} S_{q})^2 \leq \frac{ p_\alpha}{2}\,.
\end{equation}
More generally, however, one may consider a multi-sector instanton, i.e., one with charge $q$ whose support lies in multiple sectors. It is not hard to show that, in general, BPS instantons must satisfy
\begin{equation}
f_{q} S_{q} \leq \sqrt{\frac{p_{\rm tot}}{2}}  \,, ~~~~ p_{\rm tot} = \sum_{\alpha=1}^N p_\alpha \,.
\label{genfSbound}
\end{equation}
This bound will be saturated for charges of the form
\begin{equation}
q_i^{\alpha} = a  \frac{\partial}{\partial t_\alpha^i} \bras{\log\brap{\cF_{\alpha}}}\,,
\end{equation}
where $a$ is a constant.

\subsection{K\"ahler potentials in the landscape}

K\"ahler potentials that are logarithmic are very common in weakly coupled corners of the string theory landscape. Recently, they have played a central role in a series of papers that discussed EFT strings, that is, axionic strings whose core probes an infinite-distance limit in moduli space~\cite{Lanza:2020qmt,Lanza:2021udy,Grieco:2025bjy}.

A canonical example of a K\"ahler potential of the form studied here occurs for $C_4$ axions in Type IIB O3/O7 orientifold compactifications on Calabi-Yau threefolds. Here, the K\"ahler potential takes the form
\begin{equation}
\cK = - 2 \log \cV\,,
\end{equation}
where (at tree level) $\cV$ is homogeneous of degree $p = 3/2$ in the 4-cycle volumes $\tau^i$. Thus, $\cF = \cV^2$ is homogeneous of degree $p=3$, and
\begin{equation}
f_q S_q \leq \sqrt{\frac{p}{2}} = \sqrt{\frac{3}{2}}
\end{equation}
for $C_4$ axions in Type IIB O3/O7 orientifold compactifications. Saturation occurs for instantons whose action is proportional to a modulus controlling the overall volume.

Naively, $C_2$ axions have $p=6$, since the volume $\cV$ is homogeneous of degree 3 in the 2-cycle volumes $t^i$. However, there is an important subtlety here: in O3/O7 orientifold compactifications, $C_2$ axions do not reside in the same multiplets as the 2-cycle volume moduli. Said differently, the 2-cycle volumes $t^i$ are not the saxion partners of the $C_2$ axions. Indeed, the surviving 4-cycle volume moduli $\tau^i$ are even under the orientifold action, whereas the surviving $C_2$ axions are odd.

As a result, the argument used throughout this section breaks down for $C_2$ axions in O3/O7 orientifolds. (A different argument given in Appendix A of~\cite{Cicoli:2021gss} gives $fS \leq 1$ for such odd axions.) However, our argument applies straightforwardly to $C_2$ axions in O5/O9 orientifolds, as the 2-cycle volumes $t^i$ are the saxion partners of $C_2$ axions $c^i$ in such orientifolds. As discussed in, e.g., \cite{Grimm:2004uq}, the dilaton and K\"ahler moduli are mixed in O5/O9 orientifolds relative to O3/O7 orientifolds, and as a result the K\"ahler potential takes the form
\begin{equation}
\cK = - \log \cV\,,
\end{equation}
where (at tree level) $\cV$ is cubic in the 2-cycle volumes $t^i$. Consequently, we have $\cF =\cV$, $p=3$, and thus 
\begin{equation}
f_q S_q \leq \sqrt{\frac{p}{2}} = \sqrt{\frac{3}{2}}\,,
\end{equation}
as in the case of $C_4$ axions above. Saturation occurs for instantons whose action is proportional to the overall volume modulus.

By an extension of the arguments used above, it is straightforward to show directly that in $\mathcal{N}=2$ compactifications of Type IIB on Calabi-Yau threefolds, the bound $f_q S_q \leq \sqrt{3/2}$ is satisfied by D3-brane instantons, D1-brane instantons, and worldsheet instantons charged respectively under $C_4$ axions, $C_2$ axions, and $B_2$ axions.

Meanwhile, the axio-dilaton $\tau = \exp(- \phi) + i \frac{\theta}{2 \pi}$, $\theta = 2 \pi C_0$ appears in the K\"ahler potential via a term of the form
\begin{equation}
\cK = - \log(\tau + \tau^\dagger)\,,
\end{equation}
which thus has $p=1$ and
\begin{equation}
f_q S_q \leq \frac{1}{\sqrt{2}}\,.
\end{equation}
In fact, this bound is saturated by D$(-1)$-branes.
More generally, 1/2-BPS instantons in maximal supergravity have $f_q S_q = 1/\sqrt{2}$ in all dimensions $d$ \cite{Etheredge:2024amg}. This is closely related to the Sharpened Distance Conjecture~\cite{Etheredge:2022opl}. See \cite{AxionBitowers} for further discussion.

In O3/O7 orientifold compactifications, by the argument of \S\ref{ss.multiple}, instantons carrying multi-sector charge under both $C_4$ axions and the $C_0$ axion have
\begin{equation}
(f_q S_q)^2 \leq \frac{1}{2} \sum_{\alpha}  p_{\alpha} = \frac{1}{2}(3+1) =  2\,.
\end{equation}

The largest value of $p$ that we have encountered arises in the context of M-theory compactifications on $G_2$ manifolds. Here, the saxions for $C_3$ axions appear in a prepotential term with $p=7$, and hence they satisfy
\begin{equation}
f_q S_q \leq \sqrt{\frac{7}{2}}\,.
\end{equation}

The values of $p$ and $\bravv{z}_{\rm min}$ for these various examples are shown in Table \ref{tab:pzmin}. Comparing with Table \ref{tab:dual} above, we see that the values of $\bravv{z}_{\rm min}$ match in all cases where both approaches apply (namely, for $B_2$, $C_2$, and $C_4$ axions in 4d Type IIB and $C_3$ axions in 4d Type IIA). This is remarkable because, while the present approach deals exclusively with BPS instantons, the approach used in \S\ref{s.DUALITIES} applies also to directions in the instanton charge lattice outside the BPS cone. Thus, there was \emph{a priori} no reason to expect that the two bounds would agree. We speculate that this agreement may be a general result of 4d supergravity theories in the landscape, and the convex hull of instanton $z$-vectors \emph{always} contains the ball of radius $\bravv{z}_{\rm min} = \sqrt{2/p}$.  We leave this as an open question for further study.

\begin{table}
	\centering
	\renewcommand*{\arraystretch}{1.5}
	\begin{tabular}{|c|c|c|c|c|}
		\hline
		Theory & Compactification & Axion & $p$ & $\bravv{z}_{\min}$
		\\\hline
		Type IIB	& O3/O7 Orientifold	& $C_4$ & $3$ & $\sqrt{\frac23}$
		\\
		Type IIB	& O5/O9 Orientifold	& $C_2$ & $3$ & $\sqrt{\frac23}$
		\\
		Type IIB	& CY3 Manifold		& $C_0$ & $1$ & $\sqrt{2}$
		\\
		Type IIA	& CY3 Manifold		& $C_3$ & $4$ & $\sqrt{\frac12}$
		\\
		M-theory	& $G_2$ Manifold	& $C_3$ & $7$ & $\sqrt{\frac27}$
		\\\hline
	\end{tabular}
	\caption{Examples of homogeneous K\"ahler potentials $\cK$. Here, $p$ is the degree of homogeneity of $e^{-\cK}$, and $\bravv{z}_{\min}$ is the smallest charge-to-action ratio of BPS instantons charged under the listed axions in the theory. These values of $\bravv{z}_{\rm min}$ agree exactly with those of Table \ref{tab:dual} whenever both approaches apply.}
	\label{tab:pzmin}
\end{table}

\subsubsection{K\"ahler potential splitting}\label{sss.splitting}

In certain situations, a single term in the K\"ahler potential may effectively split into a sum of distinct ones, preserving the total $p_{\rm tot} = \sum_{\alpha}  p_{\alpha}$. For instance, while a generic $G_2$ compactification of M-theory has a single K\"ahler potential term with $p=7$, a compactification of M-theory on $T^7$ (or equivalently, Type II on $T^6$) yields a sum over 7 terms:
\begin{equation}
\cK = - \sum_{\alpha=1}^7 \log (T_{\alpha} + T^\dagger_{\alpha})\,,
\end{equation}
each of which has $ p_\alpha = 1$, ensuring that the sum $p_{\rm tot} = \sum_{\alpha} p_{\alpha} = 7$ is preserved. In Type IIA on a Calabi-Yau threefold with O6 planes, the dilaton and complex structure moduli together appear in a $p = 4$ term while the K\"ahler moduli appear in a $p = 3$ term~\cite{Grimm:2004ua}, again totaling the $p = 7$ degree of the M-theory limit.

Similarly, the K\"ahler potential term for $C_3$ axions in Type IIA on a Calabi-Yau threefold has $p = 4$. Under mirror symmetry, this term splits into the aforementioned terms involving $C_4$ axions and the $C_0$ axion, which have $ p_1 =3$ and $p_2=1$, respectively, ensuring that the sum $\sum_{\alpha}  p_{\alpha} = 4$ is preserved.

\subsection{A bound for axion strings}\label{ss.boundaxionstrings}

	So far we have used an homogeneous K\"ahler potential to find the bound \eqref{eq:homogeneous-kahler-awgc-bound} on $fS$ for instantons.
	While instantons are electrically charged under axions, axion strings are magnetically charged under axions.
	In this subsection, we adapt our work so far to put the bound
	\begin{equation}\label{eq:axion-string:bound}
		\frac{\cT_{\tilde q}}{\bravv{\tilde q}} \leq \sqrt{\frac p2}
	\end{equation}
	on the tension $\cT_{\tilde q}$ and magnetic charge $\bravv{\tilde q}$ of an axion string of charge ${\tilde q}$.
	
	From \eqref{eq:homogeneous-kahler-decay-constant-matrix}, the metric on axion moduli space is
	\begin{equation}
		f_{ij} = \frac{1}{2 (2 \pi)^2} \kappa_{ij}
		\,.
	\end{equation}
	Above we used this to find that the associated charge/decay constant of an instanton of charge $q$ is given by
	\begin{equation}
		\bravv{q}^2 \equiv f_q^{-2} = q_i f^{ij} q_j
		\,.
	\end{equation}
	Similarly, the (magnetic) charge of an axion string of charge $\tilde q$ is
	\begin{equation}\label{eq:axion-string:charge}
		\bravv{\tilde q}^2
		\equiv
		\brap{2\pi}^2 {\tilde q}^i f_{ij} {\tilde q}^j
		=
		\frac12 {\tilde q}^i \kappa_{ij} {\tilde q}^j
		\,.
	\end{equation}
	
	The tension, $\cT_{\tilde q}$, of a BPS axion string of magnetic charge ${\tilde q}^i$ has tension given by~\cite{Lanza:2020qmt, Lanza:2021udy} 
	\begin{equation}\label{eq:axion-string:tension}
		\cT_{\tilde q}
		= -\frac12{\tilde q}^i \partial_i \cK= \frac12{\tilde q}^i \frac{\cF_i}{\cF}
		\,.
	\end{equation}
	
	From \S\ref{ss.poslemma}, we know that $h_{ij}$ is positive semi-definite.
	Using $h_{ij}=\kappa_{ij}-\xi_{ij}$ and contracting with ${\tilde q}^i$ on both sides, we find
	\begin{equation}
		{\tilde q}^i \kappa_{ij} {\tilde q}^j - \frac1p\brap{\frac{{\tilde q}^i \cF_i}{\cF}} \geq 0
		\,.
	\end{equation}
	Using \eqref{eq:axion-string:charge}, \eqref{eq:axion-string:tension} and the definition of $\xi_{ij}$, rearranging and taking the square root, we find \eqref{eq:axion-string:bound}:
	\begin{equation}
		\frac{1}{\bravv{z^{\rm ax \, str}}} \equiv \frac{\cT_{\tilde q}}{\bravv{\tilde q}} \leq \sqrt{\frac p2}
		\,.
	\end{equation}
Note that this precisely matches the bound \eqref{fScpbound} on $\bravv{z^{\rm inst}} = \frac{1}{f_q S_q}$. We conclude that for such homogeneous K\"ahler potentials, $\bravv{z^{\rm ax \, str}}_{\rm min} = \bravv{z^{\rm inst}}_{\rm min} = \sqrt{2/p}$. 

\subsection{Bounds on the K\"ahler potential}\label{ss.Bounds}

In this subsection, we derive bounds on the degree of homogeneity $p$---and thus the minimal value $\bravv{z_{\rm min}}$---using two different approaches: (i) universal features of infinite-distance limits and (ii) bounds on imaginary traversals, as initiated in the recent works \cite{DiUbaldo:2026rly, Maldacena:2026jqd}.

\subsubsection{Infinite-distance bounds}\label{sss.infinite}

So far, we have bounded the instanton charge-to-action ratio $\bravv{z} = 1/(f_q S_q)$ in terms of the degree of homogeneity $p$. We now wish to bound this degree of homogeneity using asymptotic properties of the moduli space.

We focus our attention on the homogeneous scaling limit $t^i \rightarrow \lambda t^i$, $\lambda \rightarrow \infty$. In this limit, the kinetic term for the scaling parameter $\lambda$ takes the form
\begin{equation}
- \frac{1}{4} \kappa_{\lambda \lambda} \partial_\mu \lambda \partial^\mu \lambda  = - \frac{1}{4} \frac{\partial^2 \cK}{\partial \lambda^2} \partial_\mu \lambda \partial^\mu \lambda  = - \frac{ p}{4 \lambda^2} \partial_\mu \lambda \partial^\mu \lambda\,.
\end{equation}
By supersymmetry, the decay constant $f_{\lambda}$ of the axion partner of the scalar field $\lambda$ takes the form
\begin{equation}
f_\lambda^2 = \frac{1}{2 (2 \pi)^2} \kappa_{\lambda \lambda} \sim \frac{1}{\lambda^2}\,.
\end{equation}
The Weak Gravity Conjecture for strings \cite{ArkaniHamed:2006dz} implies that in the limit $\lambda \rightarrow \infty$, there must exist an axion string with tension bounded above as
\begin{equation}
\mathcal{T} \lesssim f_{\rm \lambda} \sim \frac{1}{\lambda} \,.
\end{equation}

Written in terms of the canonically normalized field $\hat \lambda = \sqrt{p/2} \log \lambda$, we further have
\begin{equation}
\mathcal{T} \lesssim \frac{1}{\lambda} = \exp\left(-\sqrt{\frac{2}{p}} \hat \lambda \right)\,.
\end{equation}
Next, we require the axion string scale $M_{\rm as} \equiv \sqrt{2 \pi \mathcal{T}}$ to lie above the species scale $\Lambda_{\rm QG}$, which is expected to be true of all EFT strings~\cite{Martucci:2024trp}, i.e., those associated with infinite-distance limits~\cite{Lanza:2021udy}.
According to various swampland conjectures~\cite{Lee:2019wij,Etheredge:2022opl,vandeHeisteeg:2023ubh, vandeHeisteeg:2023uxj, vandeHeisteeg:2023dlw}, this scale decays exponentially with proper field distance as 
\begin{equation}
\Lambda_{\rm QG} \sim \exp(-\alpha_{\rm QG} \hat \lambda)\,,~~~~ \frac{1}{\sqrt{(d-1)(d-2)} } \leq \alpha_{\rm QG} \leq \frac{1}{\sqrt{d-2} } \,,
\label{aQGb}
\end{equation}
where here $d=4$ is the number of spacetime dimensions.

To ensure $M_{\rm as} \gtrsim \Lambda_{\rm QG}$, we must have
\begin{equation}
  \sqrt{\frac{2}{p}} \leq \alpha_{\rm QG} \,.
\end{equation}
Together with \eqref{aQGb}, this implies a lower bound on the degree of homogeneity $p$:
\begin{equation}
p \geq 1 \,.
\label{cpbound}
\end{equation}

Placing an upper bound on the degree of homogeneity $p$ is somewhat more difficult. In general, barring the existence of a decoupling rigid field theory \cite{Marchesano:2023thx, Reece:2025zva}, we expect that the string WGC bound will be approximately saturated in an infinite-distance limit:
\begin{equation}
\cT \sim f_{\lambda}\,.
\end{equation}
 In special cases, the axion string scale $M_{\rm as} = \sqrt{2 \pi \cT}$ may further be parametrically identified with the species scale:
 \begin{equation}
 M_{\rm as} \sim \Lambda_{\rm QG} \sim \sqrt{f_{\lambda}} \sim \frac{1}{\sqrt{\lambda}} \sim \exp\left( - \sqrt{\frac{1}{2p}} \hat \lambda\right)\,.
 \end{equation}
 In these special cases, \eqref{aQGb} further implies an upper bound:
\begin{equation}
p \leq 3\,.
\end{equation}
In the following examples, we will study several examples of K\"ahler potentials that arise in known corners of the landscape. We indeed find $p \geq 1$ in all examples in the landscape with which we are familiar. Many axion sectors satisfy (and saturate) the upper bound $p \leq 3$, but others violate it.

A more general upper bound $p_{\rm tot} \leq 7$ can be obtained in the case that the limit $\lambda \rightarrow \infty$ represents a decompactification limit to 10d string theory or 11d M-theory. The Emergent String Conjecture \cite{Lee:2019wij} suggests that such a case can always be realized. We will return to this in \S\ref{ss.upperboundp} after developing the relevant background material.

In all of the examples considered above, $p$ is an integer. That $p$ is an integer between $1$ and $7$ in string theory examples was previously emphasized in~\cite{Lanza:2020qmt,Lanza:2021udy, Martucci:2024trp, Grieco:2025bjy}. In~\cite{Lanza:2021udy}, the integrality of $p$ was linked to a conjectured integral scaling relation between membrane tensions and string tensions. Using taxonomy rules for infinite-distance limits developed in \cite{Etheredge:2024tok, Etheredge:2025ahf}, one can argue that this is true more generally, as will be shown in forthcoming work~\cite{Etheredge:2026ISCproof}. Interestingly, coefficients of the form $\sqrt{2/p}$ for integer $p$ also appear in many examples in the exponential relationship between canonically normalized moduli and other quantities, such as the mass scale of the lightest tower or the potential energy; see, e.g.,~\cite{Ooguri:2006in,Grimm:2018ohb,Bedroya:2019snp,Gendler:2020dfp,Andriot:2020lea,Lanza:2021udy,Rudelius:2021oaz,Agmon:2022thq,Calderon-Infante:2022nxb,Andriot:2026lac}.

\subsubsection{A bound from imaginary traversals}\label{ss.IDBbound}

 Suppose that we have some field $\phi$ with a  kinetic term proportional to $-\lambda^\gamma \brav{\rmd \phi}^2$ for a scalar $\lambda$ with a kinetic term of the form $-\frac{p}{4\lambda^2 \kappa_d^2} \partial_\mu \lambda \partial^\mu \lambda$, and $\gamma$ a non-zero real constant.
 Consider the analytic continuation $\lambda = \lambda_0 \exp(i \alpha)$. The theory will break down when $\mathrm{Re}\,\lambda^\gamma < 0$, i.e., when $\brav{\alpha} > \pi/(2\brav{\gamma})$. The distance in field space associated with changing $\alpha$ between the extremes $-\pi/(2\brav{\gamma})$ and $+\pi/(2\brav{\gamma})$ is $\sqrt{p/2} \Delta \alpha/\kappa_d = \sqrt{p/2}\, \pi/(\kappa_d \brav{\gamma})$. If we demand that theory breaks down due to a wrong-sign kinetic term before the imaginary distance bound~\cite{DiUbaldo:2026rly,Maldacena:2026jqd} is reached, we have:
 \begin{equation}
     \sqrt{\frac{p}{2}} \frac{\pi}{\brav{\gamma} \kappa_d} \leq \pi \sqrt{\frac{d-1}{d-2}} \frac{1}{\kappa_d} \quad \Rightarrow \quad p \leq \frac{2(d-1)}{d-2} \gamma^2.
 \end{equation}
 This is somewhat speculative: the imaginary distance bound requires that some physics should invalidate the theory before a particular imaginary value of the field displacement is reached. It is not obvious that the only way for this to happen is for the kinetic term of some other field to have the wrong sign. Nonetheless, it is suggestive that we can obtain reasonable bounds, consistent with examples, from this approach.

 Let us fix $d = 4$ and consider two cases of interest. First, suppose that the field $\phi$ is an axion $\theta$ that pairs up with $\lambda$ in a chiral supermultiplet with scalar component $\lambda + i \theta$. Then the kinetic term for $\theta$ has $\gamma = -2$ (by supersymmetry), and the bound becomes
\begin{equation}
    p \leq 12.
\end{equation}
This is consistent with our examples. A more interesting bound comes from considering $\phi$ to be a gauge field $A$ with a holomorphic coupling to a scalar field $\lambda + i \theta$. Here the power of $\lambda$ is fixed to be $\gamma = 1$ by the periodicity of $\theta$.%
\footnote{
    There is an exception in special cases where $\theta \mapsto \theta + 2\pi$ is accompanied by a nontrivial monodromy on the gauge fields themselves. This occurs for Kaluza-Klein gauge theory (see, e.g.,~\cite{Heidenreich:2021yda}), which realizes $\gamma = 3$, again consistent with the bound.
} %
In this case, we obtain
\begin{equation}
    p \leq 3.
    \label{p3bound}
\end{equation}
This is nontrivially realized in many examples, as shown in Table \ref{tab:pzmin} above. In M-theory on $G_2$ manifolds, we have $p = 7$, but the modulus $\lambda$ controls a 3-cycle volume, and there is no 6-brane we can wrap on a 3-cycle to obtain 4d fields with kinetic term given directly by $\lambda$. Thus, it is unsurprising that \eqref{p3bound} is violated in these compactifications.

A more general bound on $p_{\rm tot}$ can be obtained directly from the result \eqref{sharpenedaxWGC} \cite{DiUbaldo:2026rly, Maldacena:2026jqd}, which for $d=4$ states that
\begin{equation}
f_q S_q \leq \frac{\pi}{2} \sqrt{\frac{3}{2}} \approx 1.92 M_{\rm Pl}\,.
\label{sharp4d}
\end{equation}
Setting $p_{\rm tot}=7$, our bound \eqref{fScpbound} is saturated when
\begin{equation}
f_q S_q =  \sqrt{\frac{7}{2}} \approx 1.87 M_{\rm Pl}\,,
\end{equation}
which is consistent with \eqref{sharp4d}.

In contrast, an example with $p_{\rm tot} \geq 8$ would have BPS instantons with $f_q S_q \geq 2$, in violation of \eqref{sharp4d}. It is a logical possibility that there are non-BPS instantons that have smaller action than BPS instantons and satisfy~\eqref{sharp4d}, but in the cases that we are aware of where non-BPS instantons have smaller action, they are singular solutions that cannot be uplifted to non-singular solutions in the UV completion~\cite{Bergshoeff:2004fq,Bergshoeff:2004pg}. If we assume that non-BPS instantons of lower action than BPS instantons do not arise in physically consistent settings, this again provides an argument for the upper bound $p_{\rm tot} \leq 7$.

\subsubsection{Summary of bounds on $p$}

In summary, then, swampland constraints on infinite-distance limits lead us to expect that the degrees of homogeneity $p_\alpha$, and hence also their sum $p_\text{tot}$, of a K\"ahler potential of the form \eqref{multiterm} to be positive integers.
When there is only one term in the sum, as in K\"ahler potentials of the form \eqref{Khom}, this reduces to the condition that $p$ be a positive integer.

Placing an upper bound is more difficult, but the smallest universal upper bound on $p_\text{tot}$, and hence each individual $p_\alpha$, consistent with known examples is $7$ (inclusive). We shall give a partial argument based on infinite distance for this upper bound of $7$ in \S\ref{ss.upperboundp}, once we have developed the necessary background material in the next section.

\section{A Bound on \texorpdfstring{$\alpha$}{alpha}-vectors}\label{s.ALPHA}

In a supersymmetric theory with moduli $t^i$, we may define the $\alpha$-vector of an instanton with action $S$ via
\begin{equation}
\alpha_i = \partial_i  \log S\,.
\end{equation}
The length of such an $\alpha$-vector is given by contraction with the moduli space metric $g^{ij}$:
\begin{equation}
\bravv{ \alpha }^2  = \alpha_i g^{ij} \alpha_j\,.
\end{equation}

Recall from \S\ref{ss.Supergravity} that in a supersymmetric theory, the axion and saxion form the real and complex parts of a scalar field:
\begin{equation}
T =   t - i \frac{\theta}{2 \pi} \,.
\end{equation}
This gives a relation between the kinetic terms for $\theta$ and $t$:
\begin{equation}
g_{tt} = (2 \pi)^2 f_{\theta \theta} = (2 \pi f)^2\,.
\end{equation}
A BPS instanton of charge $q = n$ has an action given by
\begin{equation}
S = 2 \pi \brav{n} t\,,
\end{equation}
so
\begin{equation}
\bravv{ \alpha}=  \sqrt{g^{tt}} \brav{\partial_t \log S} = \sqrt{g^{tt}}  \frac{2 \pi \brav{n}}{S} = \sqrt{f^{\theta \theta}} \frac{\brav{n} }{S} = \frac{\brav{n}}{fS}\,.
\end{equation}
And thus,
\begin{equation}
\frac{\brav{n}}{fS} = \bravv{ \alpha}\,.
\label{fSalphainst}
\end{equation}

This generalizes straightforwardly to supersymmetric theories with multiple axions. Set $T^i = t^i - i \frac{\theta^i}{2 \pi}$, and consider a supersymmetric instanton of charge vector $q_i$. This instanton will have action
\begin{equation}
S_q = 2 \pi \brav{q_i t^i}\,,
\end{equation}
and it will feel an effective decay constant of
\begin{equation}
f_q^2 \equiv \frac{1}{\bravv{q}^2} = \frac{1}{q_i f^{ij} q_j}\,.
\end{equation}
As discussed in \S\ref{ss.Supergravity}, the metric on axion moduli space $f_{ij}$ is related to the metric on saxion moduli space $g_{ij}$ via
\begin{equation}
f_{ij} = \frac{1}{(2 \pi)^2} g_{ij}\,.
\label{fgrel}
\end{equation}
We thus have
\begin{equation}
\bravv{\alpha}^2 = g^{ij} \partial_i(\log S) \partial_j (\log S) =  \frac{1}{S_q^2} g^{ij} (2 \pi q_i) (2 \pi q_j) = \frac{1}{S_q^2} f^{ij} q_i q_j = \frac{1}{f_q^2 S_q^2} = \bravv{z}^2\,.
\label{zequalsalpha}
\end{equation}
In fact, there are explicit BPS ``EFT instanton'' solutions that realize all of these properties, which are conveniently written in terms of the dual saxion fields~\cite{Lanza:2021udy, Martucci:2024trp}.

The upshot of this is that a bound on the magnitude of the $\alpha$-vector $\bravv{\alpha}$ of a BPS instanton immediately translates into a bound on the magnitude of the charge-to-action vector $\bravv{z}$. In this section, we derive a bound on the former.

\subsection{Bound on \texorpdfstring{$\bravv{\alpha}$}{||alpha||}}\label{ss.boundalpha}

In this subsection, we find a bound on $\bravv{\alpha}$ for BPS instantons in various string compactifications. The key observation enabling this inference is the scaling behavior of wrapped brane tensions on the volume modulus, $\rho$. In general, a $(p-1)$-brane in $d$ dimensions that descends from a wrapped $(P-1)$-brane in $D$ dimensions has a tension that scales with the canonically normalized volume modulus $\rho$ as \cite{Etheredge:2025ahf}
\begin{equation}
T_p \sim \exp\left( -\frac{p(D-2)-P(d-2)}{\sqrt{(D-d)(D-2)(d-2)}} \rho
\right)\,.
\end{equation}
For an instanton, we have $p=0$, and the ``tension'' $T_0$ is the instanton action
\begin{equation}
S \sim \exp\left(\frac{P(d-2)}{\sqrt{(D-d)(D-2)(d-2)}} \rho \right)\,.
\label{instvolscaling}
\end{equation}
In addition, a D$(p-1)$-brane in Type II string theory has a tension that scales with the canonically normalized 10d dilaton $\phi$ as
\begin{equation}
T_p \sim \exp\left( \frac{4-p}{\sqrt{8}} \phi \right)\,,
\label{Dpdilscaling}
\end{equation}
while the fundamental string tension scales as
\begin{equation}
T_2 \sim \exp\left(- \frac{1}{\sqrt{2}} \phi \right)\,.
\label{funddilscaling}
\end{equation}
As a result, the volume modulus $\rho$ and the dilaton $\phi$ contribute universally to $\vec \alpha$, with
\begin{equation}
\alpha_{ \rho} = \partial_\rho \log S\,,~~~~ \alpha_{ \phi} = \partial_\phi \log S\,.
\end{equation}
These give a lower bound on the magnitude of the $\alpha$-vector:
\begin{equation}
\bravv{\alpha}^2 \geq \alpha_{\rho}^2 + \alpha_{ \phi}^2\,.
\label{ardbound}
\end{equation}
This bound will be saturated if the instanton action depends only on the dilaton and volume modulus. In general, however, we expect additional contributions to $\bravv{\vec \alpha}$ from the ``shape moduli'' of the compactification manifold. 

As a first example, 
let us consider the $C_4$ axions $\vartheta_i$ of Type IIB string theory compactified on a Calabi-Yau threefold, which are associated with the holonomy of $C_4$ over the 4-cycles $\Sigma_4^{(i)}$. Charged instantons come from D3-branes wrapped over the 4-cycles. By \eqref{instvolscaling} actions of these instantons scale with the volume modulus as
\begin{equation}
S \sim \exp\left(\sqrt{\frac{2}{3}} \rho \right) ~~~\Rightarrow~~~ \alpha_{ \rho} = \sqrt{\frac{2}{3}} \,.
\end{equation}
The D3-brane tension is independent of the 10d dilaton $\phi$, so $\alpha_{\phi} = 0$. Thus, by \eqref{ardbound}, we see that D3-brane instantons have
\begin{equation}
\bravv{\alpha} \geq \brav{ \alpha_{\rho}} = \sqrt{\frac{2}{3}} \,.
\end{equation}
This precisely matches the minimal value $\bravv{z}_{\rm min}$ for ED3-brane instantons charged under $C_4$ axions obtained in \S\ref{s.DUALITIES} and \S\ref{s.HOMOGENEOUS} above.

\begin{table}
    \centering
    \renewcommand*{\arraystretch}{1.5}
    \begin{tabular}{|c|c|cccc|} \hline
      Theory   & Dimension & Axions & $\alpha_\phi$ & $\alpha_\rho$ & $\bravv{ \alpha_{\rm  min}}$ \\ \hline
      \multirow{ 7}{*}{Type IIB}  & \multirow{ 4}{*}{4} & $C_0$ & $\sqrt{2}$ & $0$ & $\sqrt{2}$ \\  &  & $B_2$ & $-\frac{1}{\sqrt{2}}$ & $\frac{1}{\sqrt{6}}$ & $\sqrt{\frac{2}{3}}$ \\  &  & $C_2$ & $\frac{1}{\sqrt{2}}$ & $\frac{1}{\sqrt{6}}$ & $\sqrt{\frac{2}{3}}$ \\  &  & $C_4$ & $0$ & $\sqrt{\frac{2}{3}}$ & $\sqrt{\frac{2}{3}}$ \\  \cline{2-6} & \multirow{ 3}{*}{6} & $C_0$ & $\sqrt{2}$ & $0$ & $\sqrt{2}$ \\  &  & $B_2$ & $-\frac{1}{\sqrt{2}}$ & $\frac{1}{\sqrt{2}}$ & $1$ \\  
      &  & $C_2$ & $\frac{1}{\sqrt{2}}$ & $\frac{1}{\sqrt{2}}$ & $1$ \\ \hline
      \multirow{ 2}{*}{Type IIA}  & \multirow{ 2}{*}{4} & $B_2$ & $-\frac{1}{\sqrt{2}}$ & $\frac{1}{\sqrt{6}}$ & $\sqrt{\frac{2}{3}}$ \\  &  & $C_3$ & $\sqrt{\frac{1}{8}}$ & $\sqrt{\frac{3}{8}}$ & $\frac{1}{\sqrt{2}}$ \\ 
      \hline 
      M-theory & 4 & $C_3$ & $-$ & $\sqrt{\frac{2}{7}}$ & $\sqrt{\frac{2}{7}}$ \\ \hline
    \end{tabular}
    \caption{Contributions to $\bravv{\alpha}$ from the dilaton and radion for instantons (and axion strings) charged electrically (magnetically) under various types of axions in string/M-theory compactifications. In all axion sectors where the approaches from the previous sections apply, the value of $\bravv{\alpha}_{\rm min}$ obtained here exactly matches the values of $\bravv{z}_{\rm min}$ obtained there, as shown in Tables \ref{tab:dual} and \ref{tab:pzmin}.
    }
    \label{tab:alphatypeII}
\end{table}

Next, let us consider the case of the $B_2/C_2$ axions of Type IIB string theory on a Calabi-Yau threefold, which come from the holonomies of these 2-form fields over 2-cycles. Charged instantons come from wrapped fundamental strings and wrapped D1-branes, respectively. By \eqref{instvolscaling}, these both have
\begin{equation}
S \sim \exp\left(\frac{1}{\sqrt{6}} \rho \right) ~~~\Rightarrow~~~ \alpha_{ \rho} = \frac{1}{\sqrt{6}} \,.
\end{equation}
In addition, by \eqref{Dpdilscaling} and \eqref{funddilscaling}, the F-string and D-string instantons respectively have
\begin{equation}
S \sim \exp\left(\mp \frac{1}{\sqrt{2}} \phi \right) ~~~\Rightarrow~~~ \bravv{\alpha_{\phi}} = \frac{1}{\sqrt{2}} \,.
\end{equation}
By \eqref{ardbound}, we therefore have
\begin{equation}
\bravv{\alpha}^2 \geq  \alpha_{\rho}^2 +  \alpha_{ \phi}^2 = \frac{1}{6}+ \frac{1}{2} = \frac{2}{3} \,.
\end{equation}
Once again, this precisely matches the value of $\bravv{z}_{\rm min}$ obtained through the approaches in \S\ref{s.DUALITIES} and \S\ref{s.HOMOGENEOUS} above.

This procedure may be repeated for the other types of axions/instantons in Type II compactifications on Calabi-Yau threefolds and K3 surfaces. The results are shown in Table \ref{tab:alphatypeII}. Comparing with Tables \ref{tab:dual} and \ref{tab:pzmin}, we see that the three approaches to bounding $\bravv{z}$ agree whenever two or more apply.

\begin{figure}
    \centering
    \includegraphics[width=0.7\linewidth]{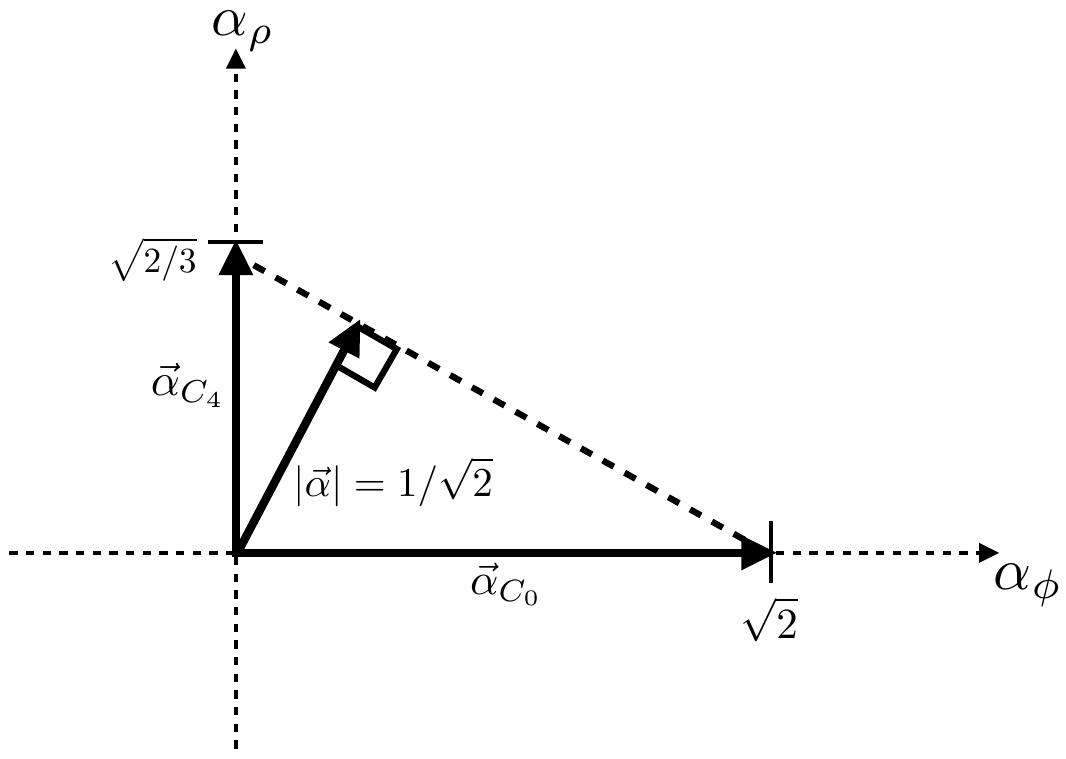}
    \caption{$\alpha$-vectors for instantons carrying charge under the $C_0$ axion and $C_4$ axions. The minimal length $\alpha$-vector corresponds to the length of the perpendicular bisector, $\bravv{\vec \alpha} = 1/\sqrt{2}$. This matches the minimal value allowed for D$2$-brane instantons in Type IIA compactifications on Calabi-Yau manifolds, as expected under mirror symmetry.}
    \label{fig:mirror}
\end{figure}

For $C_3$ axions in Type IIA string theory on a Calabi-Yau threefold, $\bravv{z} = \bravv{\alpha}$ is bounded below by $\bravv{\alpha_{\mathrm{min}}} = 1/\sqrt{2}$. This can be understood via mirror symmetry as follows: under mirror symmetry,
both $C_4$ axions and the universal $C_0$ axion of Type IIB are related to $C_3$ axions of Type IIA. From Table \ref{tab:alphatypeII}, we see that $C_4$ and $C_0$ axions have respective $\alpha$-vectors of
\begin{equation}
\vec \alpha_{C_4} = \left(0,  \sqrt{\frac{2}{3}}\right)\,,~~~~\vec \alpha_{C_0} = \left(\sqrt{2},0\right)\,.
\end{equation}
Instantons carrying multi-sector charge under both $C_0$ and $C_4$ axions will have $\alpha$-vectors that reside on the line segment between these two vectors. As shown in Figure \ref{fig:mirror}, the minimum length of an $\alpha$-vector along this line segment is $\bravv{\alpha} = 1 /\sqrt{2}$, which matches the minimal value for $C_3$ axions in Type IIA. Said differently, the minimum value of $\bravv{z} = \bravv{ \alpha} = 1/\sqrt{2}$ for D$2$-brane instantons in Type IIA compactifications corresponds, under mirror symmetry, to a diagonal direction in the instanton lattice generated by D$3$-brane instantons and D$(-1)$-brane instantons. This is the $\alpha$-vector manifestation of the prepotential splitting discussed in \S\ref{sss.splitting}.

Finally, note that the same procedure yields $\bravv{ \alpha_{ \min}} = \sqrt{2/7}$ for instantons associated with M2-branes wrapped on 3-cycles of $G_2$ manifolds, in agreement with the bound from homogeneous K\"ahler potentials in the previous section. This is the smallest value of $\bravv{z} = \bravv{\vec \alpha}$ that we have observed in the landscape. In the following subsection, we will argue that $\bravv{z} = \sqrt{2/7}$ also represents the smallest value of $\bravv{z}$ that can be obtained for BPS instantons with homogeneous K\"ahler potentials.

\subsection{Toward an upper bound on \texorpdfstring{$p$}{p}} \label{ss.upperboundp}

In \S\ref{sss.infinite}, we argued that the degree of homogeneity $p$ of a homogeneous K\"ahler potential is bounded below by $p \geq 1$. In what follows, we will argue that under certain assumptions, $p$ is bounded above by $p \leq 7$. By the argument of \S\ref{s.HOMOGENEOUS}, this in turn implies an upper bound $f_q S_q \leq \sqrt{7/2}$ on BPS instantons in such theories.

We begin with the case of a single homogeneous sector, so that the K\"ahler potential takes the form \eqref{Khom}:
\begin{equation}
\cK = -  \log(\cF(t^i))\,.
\end{equation}
We fix a starting point $t=t_\ast$, and consider the path in moduli space $t=\lambda t_\ast$, $\lambda\rightarrow\infty$. 
According to the Emergent String Conjecture, when this limit is sensible, it represents either an emergent string limit or a decompactification limit. The classical infinite distance limit may be obstructed by quantum effects~\cite{Kaufmann:2026fli,Kaufmann:2026mha,Kaufmann:2026tsy}, in which case we expect that we can consider homogeneous scaling of additional sectors of moduli to obtain a valid infinite-distance limit. The argument below generalizes straightforwardly to such a setting.

We now consider the charge
\begin{equation}
    q_i = \frac{a\lambda\cF_i}{p\cF}\Bigg|_{t=\lambda t_\ast}
    \,,
\end{equation}
for constant $a$, and assume that there is a BPS instanton of this charge.%
\footnote{
    This is not a very good assumption, since the charge $q$ is not in general properly quantized. However, one could run the same argument with the charge
    $
        q_i = a\bras{\frac{\lambda\cF_i}{p\cF}\big|_{t=\lambda t_\ast}+\eta_i\brap{a}}
    $,
    where $\eta_i\brap{a}$, a function of $a$, is the smallest number that results in $q_i$ being integrally quantized.
    Assuming that there is a BPS instanton of this charge is now more reasonable, since it is properly quantized.
    Running the argument with this charge, we find that factors of $\lambda$ cancel, so that this charge has $\bravv{z}=\sqrt{\frac2p}\brap{1+\cO\brap{\epsilon}}$ independent of $\lambda$, for
    $\epsilon\brap{a}\coloneqq\max_i\brav{\eta_i}$.
    $\epsilon$ can be made arbitrarily small by making $a$ larger, so we find that the final result, that $p\leq\frac{2}{\alpha_{\min}^2}$, still holds.
} %
Note that due to us evaluating the expression at $t=\lambda t_\ast$, and the factors of $\lambda$ canceling, this $q$ is independent of $\lambda$, so is indeed a moduli-independent quantity (in other words, it is computed in terms of a choice of $t_*$, but is then held fixed as we consider points other than $t_*$ in moduli space.) But, due to its definition, everywhere along the path $t=\lambda t_\ast$, $q_i \propto \cF_i/\cF$.
We proved in \S\ref{ss.sharpenedhom} that such a charge saturates the bound \eqref{eq:homogeneous-kahler-awgc-bound}. Using this, and that $\bravv{\alpha}=\bravv{z}$, we find that along the path $t=\lambda t_\ast$, this charge satisfies
\begin{equation}\label{2cprel}
    \sqrt{\frac2p}=\bravv{z}=\bravv{\alpha}
    \,.
\end{equation}
Hence if we can bound $\bravv{\alpha} \geq \alpha_{\min}$, then we can bound $p\leq\frac{2}{\alpha_{\min}^2}$.

If the infinite distance limit $\lambda \to \infty$ represents a decompactification limit to 10d or 11d, then in some duality frame we are dealing with a weakly-coupled compactification of string theory or M-theory. The BPS instantons in 4d must descend from BPS $(k-1)$-branes charged under $k$-form gauge fields. These branes are simply the familiar branes from string/M-theory, which were studied in \S\ref{ss.boundalpha} and are listed in Table \ref{tab:alphatypeII}. The smallest value of $\bravv{ \alpha}$ we encountered there is $\sqrt{2/7}$, corresponding to M2-branes wrapped on 3-cycles of a $G_2$ manifold.
Thus, if $\lambda \rightarrow \infty$ represents a decompactification limit to 10d or 11d, we must have $p \leq 7$. 

If $\lambda \rightarrow \infty$ instead represents a decompactification limit to $D < 10$ spacetime dimensions, we expect that there is a {\em different} infinite distance limit that we could have taken, scaling additional moduli to enlarge additional dimensions to reach $D = 10$ or $D = 11$. Even with that assumption, our argument has a loophole when applied to $p_{\mathrm{tot}}$. There may be additional moduli that were spectators to the infinite distance limit that we took, which contribute to $p_{\mathrm{tot}}$ but not to the value of $p$ that we computed. In examples that we know, this does not lead to exceptions to the bound: for instance, in Type IIB O3/O7 orientifold compactifications, the decompactification limit to 10d involves the K\"ahler moduli $T_i$ which have $p = 3$, while the complex structure moduli and axiodilaton that are spectators to this limit have $p = 4$, for $p_{\mathrm{tot}} = 7$. Our argument did not constrain the complex structure moduli, but the example still satisfies the bound. 

The expectation that $p_{\mathrm{tot}} \leq 7$ is closely related to the integral scaling conjecture of~\cite{Lanza:2021udy}, as will be discussed in the forthcoming work~\cite{Etheredge:2026ISCproof}.

\subsection{Axion strings}\label{ss.axionstring}

The relation between $\bravv{\alpha}=\bravv{z}$ extends to BPS axion strings as well. For an axion string of tension $\cT$, we define%
\footnote{
    Note that we adopt opposite sign conventions for the $\alpha$ vector of an instanton, $\alpha_i=+\partial_iS$, compared to an axion string, $\alpha_i=-\partial_i\cT$.
    This is so that the sign of the quoted values of $\alpha_\rho$ and $\alpha_\phi$ in Table \ref{tab:alphatypeII} are correct for both instantons and axion strings.
}
\begin{equation}
\alpha_i = -\partial_i \log \mathcal{T}\,,~~~~\bravv{\alpha}^2 =  \alpha_i g^{ij} \alpha_j \,.
\label{alphastring}
\end{equation}
From \eqref{eq:axion-string:charge} and \eqref{fgrel}, we have
\begin{equation}
\bravv{z}^2 = \frac{\bravv{\tilde q}^2}{\cT^2} = (2 \pi)^2 \frac{\tilde q^i f_{ij} \tilde q^j}{\cT^2} = \frac{\tilde q^i g_{ij} \tilde q^j}{\cT^2}\,.
\end{equation}
A BPS axion string satisfies
\begin{equation}
\cT =  -\frac{1}{2}\tilde q^i \partial_i \cK\,,
\end{equation}
hence
\begin{equation}
\partial_i \cT = - \frac{1}{2} \tilde q^j \partial_i \partial_j \cK = - \frac{1}{2}\tilde q^j \kappa_{ij} = - \tilde q^j g_{ij}\,.
\end{equation}
So
\begin{equation}
\bravv{\alpha}^2 = \frac{ \tilde q^i g_{ij} g^{jk} g_{kl} \tilde q^j}{\cT^2} = \frac{ \tilde q^i g_{ij} \tilde q^j}{\cT^2}\,.
\end{equation}
Hence,
\begin{equation}
\bravv{\alpha}^2 =  \bravv{z}^2\,.
\end{equation}

In Type IIB string theory in 4d, axion strings charged magnetically under $C_0$, $B_2$, $C_2$, and $C_4$ axions come from wrapped D7-branes, NS5-branes, D5-branes, and D3-branes, respectively. For Type IIA string theory in 4d, axion strings charged magnetically under $C_3$ axions come from D4-branes wrapped on 3-cycles. For M-theory on a $G_2$ manifold, axion strings charged magnetically under $C_3$ axions come from M5-branes wrapped on 4-cycles. With the sign convention used in \eqref{alphastring}, the values of $\alpha_\rho$ and $\alpha_\phi$ obtained for each of these magnetically charged axion strings precisely match the values of $\alpha_\rho$ and $\alpha_\phi$ obtained for the corresponding electrically charged instantons. Consequently, the value of $\bravv{\alpha}_{\rm min} = \sqrt{\alpha_\rho^2 +\alpha_\phi^2}$ for the axion strings precisely matches that of the corresponding instantons, so
Table \ref{tab:alphatypeII} applies equally well to both instantons and axion strings.

Putting these results together, we have
\begin{equation}
\bravv{z^{\rm ax \, str}}_{\rm min} = \bravv{\alpha^{\rm ax \, str}}_{\rm min} = \bravv{\alpha^{\rm inst}}_{\rm min} = \bravv{z^{\rm inst}}_{\rm min} \,,
\end{equation}
which fits perfectly with the result
\begin{equation}
\bravv{z^{\rm ax \, str}}_{\rm min} = \bravv{z^{\rm inst}}_{\rm min} = \sqrt{\frac{2}{p}}
\end{equation}
obtained for homogeneous K\"ahler potentials in \S\ref{ss.sharpenedhom} and \S\ref{ss.boundaxionstrings}.

\section{Axion Inflation}\label{s.INFLATION}

In this section, we comment briefly on the implications of our work for axion inflation.

Natural inflation was first proposed in \cite{Freese:1990rb} as a candidate mechanism for generating super-Planckian inflaton field ranges. The key idea is that the discrete axion shift symmetry protects the axion potential from problematic Planck-suppressed operators, which would destroy the flatness of the potential.

However, super-Planckian natural inflation requires super-Planckian decay constants, which are difficult to construct in string theory \cite{Banks:2003sx}. This can be understood from the Weak Gravity Conjecture, as observed in the original paper \cite{ArkaniHamed:2006dz}. For a natural inflation potential of the form
\begin{equation}
V(\theta) = \Lambda_{\rm UV}^4 e^{-S} \left( 1 - \cos \theta \right) + O(e^{-2S})\,,
\end{equation}
natural inflation requires $f \gtrsim 5 M_{\rm Pl}$, while control of the instanton expansion typically requires $S \gg 1$. As we have seen, the axion WGC requires $f S \lesssim M_{\rm Pl}$, which is in tension with this bound.

In response to these constraints, several ideas were proposed for generating an effective super-Planckian decay constant in string theory. The first, known as $N$-flation \cite{liddle:1998jc, dimopoulos:2005ac}, involves recruiting $N$ axions with order-one decay constants, then traversing the $N$-dimensional space diagonal in axion moduli space to generate an effective decay constant of $O(\sqrt{N})$. The second, known as alignment \cite{Kim:2004rp}, uses mixing between two or more axions to generate a large effective decay constant. This can either take the form of kinetic mixing or of mass mixing; here, we will focus on the former.

In general, these multi-axion models also run into trouble with the WGC, as discussed in \cite{Rudelius:2014wla, Rudelius:2015xta, Montero:2015ofa, Brown:2015iha, Heidenreich:2015wga}. Surveys of the string landscape have also failed to find successful examples of $N$-flation or axion alignment (see, e.g.,~\cite{Bachlechner:2015qja, Brown:2015lia,Junghans:2015hba, Palti:2015xra,  Conlon:2016aea, long:2016jvd}). In this section, we revisit $N$-flation and kinetic alignment in light of the precise bounds derived in the present work. In short, we will argue that simple, isotropic models of $N$-flation are ruled out by our bounds, while some models of kinetic alignment survive.

\subsection{Bounds on \texorpdfstring{$N$}{N}-flation}\label{ss.Nflation}

Consider a theory with $N$ axions, and take the axion kinetic matrix to be given simply by
\begin{equation}
\kappa_{ij} = f^2 \delta_{ij}\,.
\end{equation}
Then, the effective decay constant is given by
\begin{equation}
f_{\rm eff} = \sqrt{N} f\,,
\end{equation}
which is the length of the space diagonal of an $N$-dimensional hypercube of side length $f$.

Consider the instanton of charge $q_i = (1, 1, ..., 1)$. This instanton has action $S_{q} =  q_i S^i = \sum_i S^i > N$, since perturbative control requires $S^i >1$ for all $i$. The charge of the instanton is given by
\begin{equation}
\bravv{q} = \sqrt{q_i \kappa^{ij}q_j} = \frac{\sqrt{N}}{f}\,.
\end{equation}
The BPS axion WGC derived in \eqref{fScpbound} then gives
\begin{equation}
\frac{1}{\sqrt{N}f} \geq \frac{\bravv{q}}{S_q} \geq \sqrt{\frac{2}{p}}\,.
\end{equation}
This yields
\begin{equation}
f_{\rm eff} \leq \sqrt{\frac{p}{2}}\,,
\end{equation}
which in particular is independent of $N$. Using multiple axion sectors, the degree of homogeneity $p$ can be replaced by the sum $p_{\rm tot}$, as in \eqref{fScpbound}.  
However, we have argued above that $p_{\rm tot} \leq 7$, which precludes parametrically super-Planckian decay constants.

\subsection{Bounds on Kinetic Alignment}\label{ss.Alignment}

Axion alignment is much more subtle, and in general, it is not ruled out by the BPS axion WGC. However, certain types of alignment are constrained, as we explain below.

Consider a theory with two axions with axion kinetic matrix
\begin{equation}
\kappa_{ij} = f^2 \left(
\begin{array}{cc}
1 & 1- \epsilon \\
1- \epsilon & 1
\end{array}
\right)\,.
\end{equation}
Kinetic alignment occurs in the limit $\epsilon \rightarrow 0$.

Let us suppose that the BPS cone is generated by two instantons of respective charges $q_i^{(1)} = (1,0)$, $q_i^{(2)}=(0,1)$.
The diameter $\mathcal{D} = 2 \pi f_{\rm eff}$ of the axion fundamental domain is given by the length of the diagonal between $\theta_1 = \theta_2 = -\pi$ and $\theta_1 = \theta_2 = + \pi$, which is equal to 
\begin{equation}
\mathcal{D} =2 \pi f_{\rm eff} = \int_{- \pi}^\pi d t \sqrt{\kappa_{ij} \dot \theta^i \dot \theta^j} = 4 \pi f \sqrt{1 - \frac{\epsilon}{2}}\,.
\end{equation}
Let us now suppose that the instanton of charge $q_i=(1, 1)$ has action given by $S_{1,1} = S_1 + S_2 \geq 2$ and satisfies the axion WGC bound \eqref{fScpbound}. We then have
\begin{equation}
\sqrt{\frac{2}{p}} \leq \bravv{z} \leq \frac{\sqrt{q_i \kappa^{ij} q_j}}{2} = \frac{1}{2 f \sqrt{1 - \frac{\epsilon}{2}}} = \frac{1}{f_{\rm eff}}\,.
\end{equation}
We see that in this scenario, the effective decay constant is bounded above as $f_{\rm eff} \leq \sqrt{p/2}$, so alignment fails to produce a super-Planckian traversal.

However, the story is different if we instead consider a kinetic matrix of the form
\begin{equation}
\kappa_{ij} = f^2 \left(
\begin{array}{cc}
1 & \epsilon -1 \\
 \epsilon -1 & 1
\end{array}
\right)\,.
\end{equation}
Once again, alignment occurs in the limit $\epsilon \rightarrow 0$.

Let us again suppose that the BPS cone is generated by two instantons of respective charges $q_i^{(1)} = (1,0)$, $q_i^{(2)}=(0,1)$.
The diameter $\mathcal{D} = 2 \pi f_{\rm eff}$ of the axion fundamental domain is now given by the length of the diagonal between $\theta_1 = -\theta_2 = -\pi$ and $\theta_1 = -\theta_2 = + \pi$, which is again equal to 
\begin{equation}
\mathcal{D} =2 \pi f_{\rm eff} = \int_{- \pi}^\pi d t \sqrt{\kappa_{ij} \dot \theta^i \dot \theta^j} = 4 \pi f \sqrt{1 - \frac{\epsilon}{2}}\,.
\end{equation}
Let us see what happens when we apply our bound \eqref{fScpbound} to the instanton of charge $q_i=(1, 1)$ with action $S_{(1,1)} = S_1+ S_2 \geq 2$. This gives
\begin{equation}
\sqrt{\frac{2}{p}} \leq \bravv{z} \leq \frac{\sqrt{q_i \kappa^{ij} q_j}}{2} = \frac{1}{ f \sqrt{2\epsilon}}\,.
\end{equation}
Taking $\epsilon f^2 \rightarrow 0$, $f \rightarrow \infty$, we find that this bound can be satisfied even though $f_{\rm eff}$ can be parametrically large: the BPS axion WGC, applied to the instanton of charge $q_i = (1, 1)$, does not usefully restrict axion alignment.

\begin{figure}
    \centering
\includegraphics[width=0.45\linewidth]{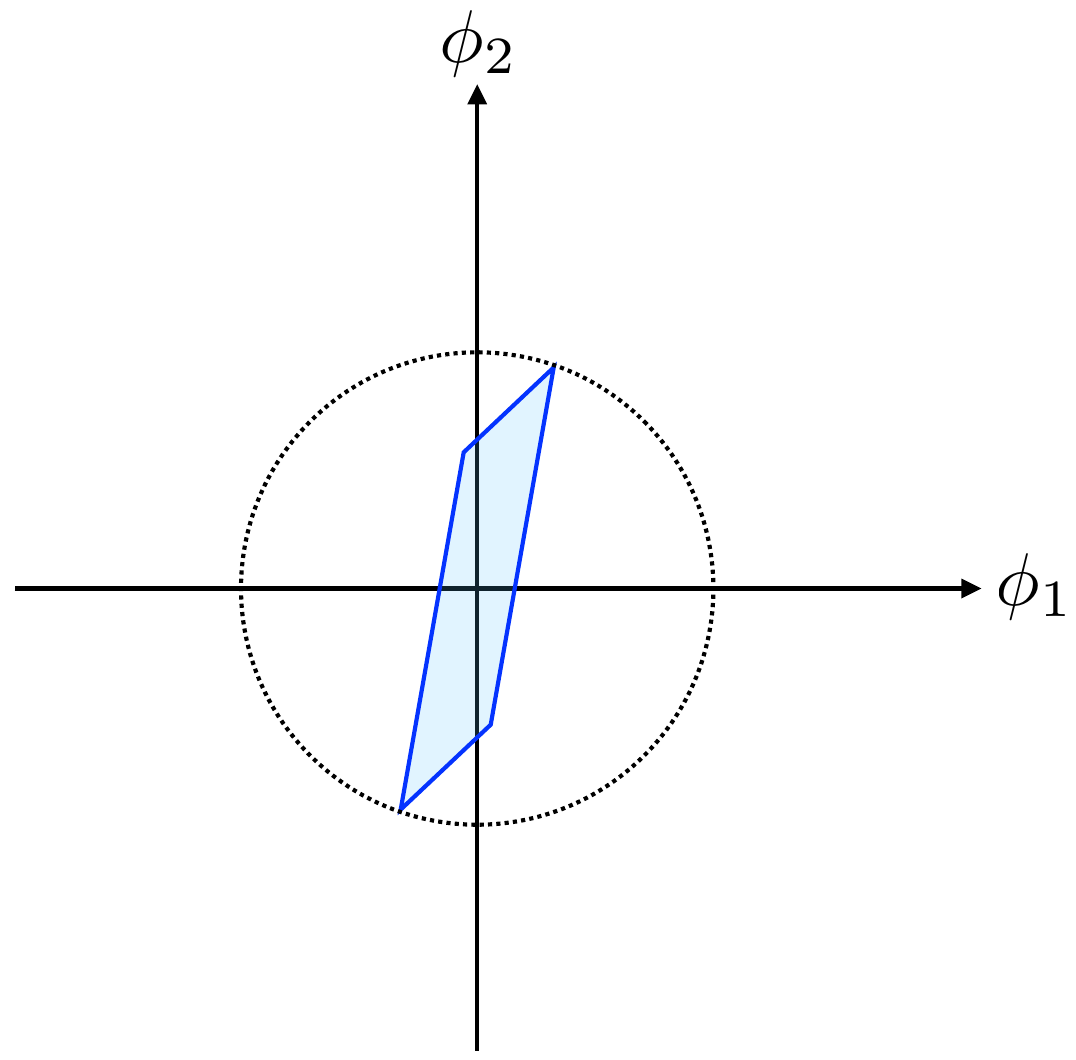}
\includegraphics[width=0.45\linewidth]{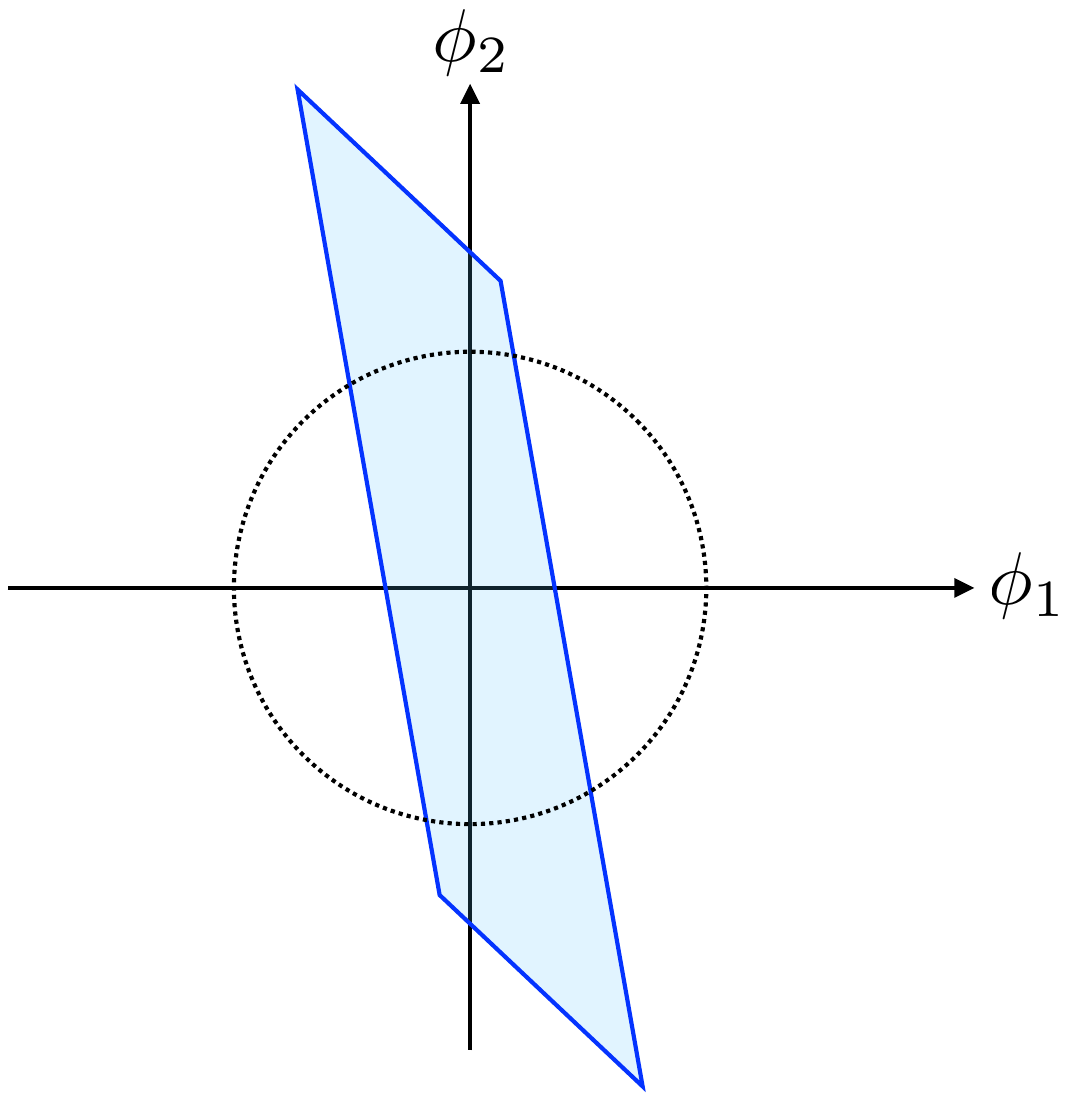}
    \caption{One orientation of alignment is incompatible with the BPS axion WGC (left), but the other is not (right). Here, the blue shaded region represents the axion fundamental domain, while the dashed black circle has radius $r = \pi \sqrt{p/2}$. The axes $\phi_1$ and $\phi_2$ depict canonically normalized axions.}
\label{fig:alignmentvsanti}
\end{figure}

To place a meaningful bound, we would instead need to consider the instanton of charge $q_i = (1, -1)$. This instanton has
\begin{equation}
\bravv{z} = \frac{\sqrt{q_i \kappa^{ij} q_j}}{S_{(1,-1)}} = \frac{2}{f_{\rm eff} S_{(1, -1)}}\,.
\end{equation}
From here, the BPS axion WGC would yield the familiar result $f_{\rm eff} \leq \sqrt{p/2}$ if we could argue $S_{(1, -1)} \geq 2$, as we did in the previous example above. However, the charge $(1, -1)$ lies outside the BPS cone, so we cannot necessarily apply the BPS axion WGC to it, nor can we simply state that the instanton action is given by the sum of the instanton actions for the generators of the BPS cone.

We conclude that one orientation of kinetic alignment is ruled out by the BPS axion WGC, while the other is not. The difference between these two is depicted in Figure \ref{fig:alignmentvsanti}. This is related to the observation of \cite{Rudelius:2015xta}, which rigorously ruled out decay constant ``anti-alignment'' of $C_4$ axions in Type IIB compactifications but could not rule out alignment in the opposite orientation.

This result highlights an important lesson: the approaches utilized in \S\ref{s.HOMOGENEOUS} and \S\ref{s.ALPHA} of this paper do not imply a convex hull condition on instanton charge-to-action vectors, as they do not apply to instantons outside the BPS cone. In contrast, the approach of \S\ref{s.DUALITIES} does imply a convex hull condition in the situations (when it applies), as does the sharpened axion WGC \eqref{sharpenedaxWGC}. These bounds do rule out the models of kinetic alignment presented in this subsection.

However, it must be noted that none of the axion WGC bounds presented here--nor those derived in \cite{DiUbaldo:2026rly, Maldacena:2026jqd}---have definitively ruled out natural inflation. The ``extra instanton loophole'' \cite{Bachlechner:2015qja, Brown:2015lia, Hebecker:2015rya, Heidenreich:2015wga} and the ``small action loophole'' \cite{delaFuente:2014aca} remain open,%
\footnote{
    See \S7.1.5 of \cite{Harlow:2022ich} for a review.
} %
and until they are closed or threaded by means of an explicit model, the question of natural inflation in quantum gravity cannot be answered definitively.

\section{Discussion}\label{s.DISC}

In this work, we have used three different approaches to derive precise statements of the axion WGC in supersymmetric quantum gravity theories. The results obtained are complementary in the sense that whenever two or more approaches apply, they yield the same results. All of the bounds we obtain are consistent with---and slightly stronger than---the sharpened axion WGC bound \eqref{sharpenedaxWGC} obtained in the recent works \cite{DiUbaldo:2026rly, Maldacena:2026jqd}.

It must be emphasized that the agreement between the three approaches was not guaranteed \emph{a priori}. In particular, while our second and third approaches (involving homogeneous K\"ahler potentials and instanton $\alpha$-vectors, respectively) dealt specifically with BPS instantons, the first approach instead used the string duality web in combination with higher-form versions of the WGC to establish a convex hull condition that (conjecturally) applies both inside and outside the cone of BPS instantons. The fact that our approaches yield identical bounds raises the interesting possibility that the bounds on BPS instantons derived in \S\ref{s.HOMOGENEOUS} and \S\ref{s.ALPHA} can always be extended to convex hull conditions over the entirety of the instanton charge lattice. Further work in this direction is needed.

Our work suggests that BPS instantons in 4d supergravity theories must satisfy a bound of the form
\begin{equation}
f_q S_q \leq \sqrt{\frac{p}{2}} M_{\rm Pl}\,,
\end{equation}
where $p$ is an integer between 1 and 7, inclusive. Within the string/M-theory landscape, there are several especially noteworthy values of $p$. 1/2-BPS instantons in maximal supergravity (as well as D$(-1)$-brane instantons charged under the universal axion $C_0$ in Type IIB compactifications) have $p=1$ and saturate the bound \eqref{fScpbound}. $C_4$ axions in O3/O7-orientifolds and $C_2$ axions in O5/O9 orientifolds have $p=3$, which is also the value associated with instantons coming from particles in pure 5d supergravity wrapped on a circle. The largest value observed, $p=7$, corresponds to $C_3$ axions in M-theory compactified on $G_2$ manifolds. In \S\ref{ss.IDBbound}, we have given a suggestive (but not rigorous) argument that the imaginary distance bound may require the stronger condition $p \leq 3$ for any axion that is part of the modulus field determining a gauge coupling.
Our work also suggests that BPS axion strings in 4d supergravity theories obey a bound of the form
\begin{equation}
\frac{\cT_{\tilde q}}{\bravv{\tilde q}} \leq \sqrt{\frac{p}{2}} M_{\rm Pl}\,,
\end{equation}
for the same value of $p$.

In this work, we have largely ignored the topic of gravitational instantons. However, it is interesting to note that in axion-dilaton-gravity, wormhole solutions do not exist above the critical value
\begin{equation}
\bravv{\alpha} = \sqrt{\frac{d-2}{d-1}}\,,
\end{equation}
in $d$ spacetime dimensions.
This corresponds to the value $\bravv{\alpha} = \sqrt{2/3}$ in 4d, which in turn corresponds to the bound $p \leq 3$ on homogeneous K\"ahler potentials studied in \S\ref{s.HOMOGENEOUS}. In contrast, for $\bravv{\alpha} < \sqrt{2/3}$, gravitational wormholes do not merely exist, but they also have a \emph{larger} charge-to-action ratio $\bravv{z} = \bravv{\alpha}$ than do the BPS instantons, as observed in \cite{Bergshoeff:2004fq,Bergshoeff:2004pg,Arkani-Hamed:2007cpn}. We still do not have an explanation for this strange phenomenon.

We observed a phenomenon of superpotential splitting, in which a single homogeneous term of the K\"ahler potential may split into a sum of multiple homogeneous terms in such a way that the sum of the degrees of homogeneity $p_{\rm tot}=\sum_\alpha p_\alpha$ is preserved. This splitting manifested as a diagonal direction in the convex hull of $\alpha$-vectors. It would be worthwhile to further explore this phenomenon and its consequences.

Finally, we note that while~\cite{DiUbaldo:2026rly,Maldacena:2026jqd} have placed the axion WGC on a sounder footing, they have also raised new questions and provided new tools that should be applied broadly. Is an instanton obeying the axion WGC the only cure for the pathologies associated with Giddings-Strominger wormholes, and if not, do the alternatives have interesting physical implications of their own? Are there other applications of the gravitational path integral to axion physics? These questions may shed further light on the Strong CP problem, dark matter, inflation, and dark energy.

\section*{Acknowledgements}

We are grateful for conversations with David Andriot, Christian Aoufia, Josh Benabou, Michele Cicoli, Bernardo Fraiman, Naomi Gendler, Ben Heidenreich, Elias Kiritsis, Dieter L\"ust, Liam McAllister, Luca Melotti, Sanjay Raman, Alex Stewart, and Cynthia Yan.
MR is supported in part by the DOE Grant DE-SC0013607.
The work of TR was supported in part by STFC through grant ST/T000708/1 and by the Royal Society through grant RGS/R2/252603.
The work of CT was supported by a studentship from STFC through grant ST/Y509334/1 and by the Royal Society through grant RGS/R2/252603.

\bibliographystyle{JHEP}
\bibliography{ref}

@article{Grieco:2025bjy,
    author = "Grieco, Alessandra and Ruiz, Ignacio and Valenzuela, Irene",
    title = "{EFT strings and dualities in 4d $\mathcal{N}=1$}",
    eprint = "2504.16984",
    archivePrefix = "arXiv",
    primaryClass = "hep-th",
    reportNumber = "IFT-UAM/CSIC-25-39, CERN-TH-2025-086",
    month = "4",
    year = "2025"
}

@article{Lanza:2021udy,
    author = "Lanza, Stefano and Marchesano, Fernando and Martucci, Luca and Valenzuela, Irene",
    title = "{The EFT stringy viewpoint on large distances}",
    eprint = "2104.05726",
    archivePrefix = "arXiv",
    primaryClass = "hep-th",
    doi = "10.1007/JHEP09(2021)197",
    journal = "JHEP",
    volume = "09",
    pages = "197",
    year = "2021"
}

@article{Lanza:2020qmt,
    author = "Lanza, Stefano and Marchesano, Fernando and Martucci, Luca and Valenzuela, Irene",
    title = "{Swampland Conjectures for Strings and Membranes}",
    eprint = "2006.15154",
    archivePrefix = "arXiv",
    primaryClass = "hep-th",
    doi = "10.1007/JHEP02(2021)006",
    journal = "JHEP",
    volume = "02",
    pages = "006",
    year = "2021"
}

@article{Castellano:2023stg,
    author = "Castellano, Alberto and Ruiz, Ignacio and Valenzuela, Irene",
    title = "{Universal Pattern in Quantum Gravity at Infinite Distance}",
    eprint = "2311.01501",
    archivePrefix = "arXiv",
    primaryClass = "hep-th",
    reportNumber = "CERN-TH-2023-203",
    doi = "10.1103/PhysRevLett.132.181601",
    journal = "Phys. Rev. Lett.",
    volume = "132",
    number = "18",
    pages = "181601",
    year = "2024"
}

@article{Castellano:2023jjt,
    author = "Castellano, Alberto and Ruiz, Ignacio and Valenzuela, Irene",
    title = "{Stringy evidence for a universal pattern at infinite distance}",
    eprint = "2311.01536",
    archivePrefix = "arXiv",
    primaryClass = "hep-th",
    reportNumber = "CERN-TH-2023-204",
    doi = "10.1007/JHEP06(2024)037",
    journal = "JHEP",
    volume = "06",
    pages = "037",
    year = "2024"
}

@article{Etheredge:2023odp,
    author = "Etheredge, Muldrow and Heidenreich, Ben and McNamara, Jacob and Rudelius, Tom and Ruiz, Ignacio and Valenzuela, Irene",
    title = "{Running decompactification, sliding towers, and the distance conjecture}",
    eprint = "2306.16440",
    archivePrefix = "arXiv",
    primaryClass = "hep-th",
    reportNumber = "ACFI-T23-02, CERN-TH-2023-121, IFT-UAM/CSIC-23-78",
    doi = "10.1007/JHEP12(2023)182",
    journal = "JHEP",
    volume = "12",
    pages = "182",
    year = "2023"
}

@article{Benakli:2020pkm,
    author = {Benakli, Karim and Branchina, Carlo and Lafforgue-Marmet, Ga{\"e}tan},
    title = "{Revisiting the scalar weak gravity conjecture}",
    eprint = "2004.12476",
    archivePrefix = "arXiv",
    primaryClass = "hep-th",
    doi = "10.1140/epjc/s10052-020-8268-0",
    journal = "Eur. Phys. J. C",
    volume = "80",
    number = "8",
    pages = "742",
    year = "2020"
}

@misc{Etheredge:2026ISCproof,
	Author = {Etheredge, Muldrow and Herraez, Alvaro and Melotti, Luca and Luest, Dieter},
	Note = {to appear},
	Title = {Proving the integral scaling conjecture with Brane Taxonomy},
	Year = {2026}}

@article{Palti:2015xra,
    author = "Palti, Eran",
    title = "{On Natural Inflation and Moduli Stabilisation in String Theory}",
    eprint = "1508.00009",
    archivePrefix = "arXiv",
    primaryClass = "hep-th",
    doi = "10.1007/JHEP10(2015)188",
    journal = "JHEP",
    volume = "10",
    pages = "188",
    year = "2015"
}

@article{Junghans:2015hba,
    author = "Junghans, Daniel",
    title = "{Large-Field Inflation with Multiple Axions and the Weak Gravity Conjecture}",
    eprint = "1504.03566",
    archivePrefix = "arXiv",
    primaryClass = "hep-th",
    reportNumber = "LMU-ASC-21-15",
    doi = "10.1007/JHEP02(2016)128",
    journal = "JHEP",
    volume = "02",
    pages = "128",
    year = "2016"
}

@misc{AxionBitowers,
	Author = {Etheredge, Muldrow and Heidenreich, Ben and Reece, Matt and Rudelius, Tom},
	Note = {to appear},
	Title = {Axion Bitowers in Quantum Gravity}}

@article{Lee:2018spm,
    author = "Lee, Seung-Joo and Lerche, Wolfgang and Weigand, Timo",
    title = "{A Stringy Test of the Scalar Weak Gravity Conjecture}",
    eprint = "1810.05169",
    archivePrefix = "arXiv",
    primaryClass = "hep-th",
    reportNumber = "CERN-TH-2018-220",
    doi = "10.1016/j.nuclphysb.2018.11.001",
    journal = "Nucl. Phys. B",
    volume = "938",
    pages = "321--350",
    year = "2019"
}

@article{Etheredge:2025ahf,
    author = "Etheredge, Muldrow",
    title = "{Taxonomy of branes in infinite distance limits}",
    eprint = "2505.10615",
    archivePrefix = "arXiv",
    primaryClass = "hep-th",
    reportNumber = "ACFI-T25-02",
    doi = "10.1007/JHEP10(2025)200",
    journal = "JHEP",
    volume = "10",
    pages = "200",
    year = "2025"
}

@article{Rudelius:2022gyu,
    author = "Rudelius, Tom",
    title = "{Constraints on early dark energy from the axion weak gravity conjecture}",
    eprint = "2203.05575",
    archivePrefix = "arXiv",
    primaryClass = "hep-th",
    doi = "10.1088/1475-7516/2023/01/014",
    journal = "JCAP",
    volume = "01",
    pages = "014",
    year = "2023"
}

@article{Marchesano:2023thx,
    author = "Marchesano, Fernando and Melotti, Luca and Paoloni, Lorenzo",
    title = "{On the moduli space curvature at infinity}",
    eprint = "2311.07979",
    archivePrefix = "arXiv",
    primaryClass = "hep-th",
    doi = "10.1007/JHEP02(2024)103",
    journal = "JHEP",
    volume = "02",
    pages = "103",
    year = "2024"
}

@article{Maldacena:2026jqd,
    author = "Maldacena, Juan and Maloney, Alexander and McPeak, Brian",
    title = "{Wormholes and the imaginary distance bound}",
    eprint = "2605.05336",
    archivePrefix = "arXiv",
    primaryClass = "hep-th",
    month = "5",
    year = "2026"
}

@article{Rudelius:2024mhq,
    author = "Rudelius, Tom",
    title = "{An Introduction to the Weak Gravity Conjecture}",
    eprint = "2409.02161",
    archivePrefix = "arXiv",
    primaryClass = "hep-th",
    doi = "10.1080/00107514.2024.2391206",
    journal = "Contemp. Phys.",
    volume = "1",
    pages = "14",
    year = "2024"
}

@article{DiUbaldo:2026rly,
    author = "Di Ubaldo, Gabriele and Iliesiu, Luca V. and Lin, Henry W. and Yan, Cynthia",
    title = "{Positivity of the gravitational path integral implies the axionic weak gravity conjecture}",
    eprint = "2605.05305",
    archivePrefix = "arXiv",
    primaryClass = "hep-th",
    reportNumber = "RIKEN-iTHEMS-Report-26",
    month = "5",
    year = "2026"
}

@article{Grimm:2004uq,
    author = "Grimm, Thomas W. and Louis, Jan",
    title = "{The Effective action of N = 1 Calabi-Yau orientifolds}",
    eprint = "hep-th/0403067",
    archivePrefix = "arXiv",
    reportNumber = "LPTENS-04-14",
    doi = "10.1016/j.nuclphysb.2004.08.005",
    journal = "Nucl. Phys. B",
    volume = "699",
    pages = "387--426",
    year = "2004"
}

@article{Kim:2004rp,
    author = "Kim, Jihn E. and Nilles, Hans Peter and Peloso, Marco",
    title = "{Completing natural inflation}",
    eprint = "hep-ph/0409138",
    archivePrefix = "arXiv",
    doi = "10.1088/1475-7516/2005/01/005",
    journal = "JCAP",
    volume = "01",
    pages = "005",
    year = "2005"
}

@article{Etheredge:2023zjk,
    author = "Etheredge, Muldrow and Heidenreich, Ben",
    title = "{Geodesic gradient flows in moduli space}",
    eprint = "2311.18693",
    archivePrefix = "arXiv",
    primaryClass = "hep-th",
    reportNumber = "AFCI-T23-09",
    doi = "10.1007/JHEP03(2025)035",
    journal = "JHEP",
    volume = "03",
    pages = "035",
    year = "2025"
}

@article{Etheredge:2023usk,
    author = "Etheredge, Muldrow",
    title = "{Dense geodesics, tower alignment, and the Sharpened Distance Conjecture}",
    eprint = "2308.01331",
    archivePrefix = "arXiv",
    primaryClass = "hep-th",
    reportNumber = "ACFI-T23-04",
    doi = "10.1007/JHEP01(2024)122",
    journal = "JHEP",
    volume = "01",
    pages = "122",
    year = "2024"
}

@article{Calderon-Infante:2020dhm,
	Archiveprefix = {arXiv},
	Author = {Calder\'on-Infante, Jos\'e and Uranga, Angel M. and Valenzuela, Irene},
	Doi = {10.1007/JHEP03(2021)299},
	Eprint = {2012.00034},
	Journal = {JHEP},
	Pages = {299},
	Primaryclass = {hep-th},
	Reportnumber = {IFT-UAM/CSIC-20-169},
	Title = {{The Convex Hull Swampland Distance Conjecture and Bounds on Non-geodesics}},
	Volume = {03},
	Year = {2021}}

@article{vandeHeisteeg:2023ubh,
    author = "van de Heisteeg, Damian and Vafa, Cumrun and Wiesner, Max",
    title = "{Bounds on Species Scale and the Distance Conjecture}",
    eprint = "2303.13580",
    archivePrefix = "arXiv",
    primaryClass = "hep-th",
    doi = "10.1002/prop.202300143",
    journal = "Fortsch. Phys.",
    volume = "71",
    number = "10-11",
    pages = "2300143",
    year = "2023"
}

@article{Etheredge:2024tok,
    author = "Etheredge, Muldrow and Heidenreich, Ben and Rudelius, Tom and Ruiz, Ignacio and Valenzuela, Irene",
    title = "{Taxonomy of infinite distance limits}",
    eprint = "2405.20332",
    archivePrefix = "arXiv",
    primaryClass = "hep-th",
    reportNumber = "ACFI-T24-04, CERN-TH-2024-067, IFT-UAM/CSIC-23-64",
    doi = "10.1007/JHEP03(2025)213",
    journal = "JHEP",
    volume = "03",
    pages = "213",
    year = "2025"
}

@article{Etheredge:2024amg,
    author = "Etheredge, Muldrow and Heidenreich, Ben and Rudelius, Tom",
    title = "{A Distance Conjecture for branes}",
    eprint = "2407.20316",
    archivePrefix = "arXiv",
    primaryClass = "hep-th",
    reportNumber = "ACFI-T24-05",
    doi = "10.1007/JHEP09(2025)155",
    journal = "JHEP",
    volume = "09",
    pages = "155",
    year = "2025"
}

@article{Marchesano:2019ifh,
	Archiveprefix = {arXiv},
	Author = {Marchesano, Fernando and Wiesner, Max},
	Doi = {10.1007/JHEP08(2019)088},
	Eprint = {1904.04848},
	Journal = {JHEP},
	Pages = {088},
	Primaryclass = {hep-th},
	Reportnumber = {IFT-UAM/CSIC-19-049},
	Title = {{Instantons and infinite distances}},
	Volume = {08},
	Year = {2019}}

@article{Etheredge:2022opl,
    author = "Etheredge, Muldrow and Heidenreich, Ben and Kaya, Sami and Qiu, Yue and Rudelius, Tom",
    title = "{Sharpening the Distance Conjecture in diverse dimensions}",
    eprint = "2206.04063",
    archivePrefix = "arXiv",
    primaryClass = "hep-th",
    reportNumber = "ACFI-T22-07",
    doi = "10.1007/JHEP12(2022)114",
    journal = "JHEP",
    volume = "12",
    pages = "114",
    year = "2022"
}

@article{Heidenreich:2020upe,
	Archiveprefix = {arXiv},
	Author = {Heidenreich, Ben},
	Doi = {10.1007/JHEP11(2020)029},
	Eprint = {2006.09378},
	Journal = {JHEP},
	Pages = {029},
	Primaryclass = {hep-th},
	Reportnumber = {ACFI-T20-07},
	Title = {{Black Holes, Moduli, and Long-Range Forces}},
	Volume = {11},
	Year = {2020}}

@article{Hebecker:2018ofv,
	Archiveprefix = {arXiv},
	Author = {Hebecker, Arthur and Mikhail, Thomas and Soler, Pablo},
	Doi = {10.3389/fspas.2018.00035},
	Eprint = {1807.00824},
	Journal = {Front. Astron. Space Sci.},
	Pages = {35},
	Primaryclass = {hep-th},
	Title = {{Euclidean wormholes, baby universes, and their impact on particle physics and cosmology}},
	Volume = {5},
	Year = {2018}}

@article{Hebecker:2016dsw,
	Archiveprefix = {arXiv},
	Author = {Hebecker, Arthur and Mangat, Patrick and Theisen, Stefan and Witkowski, Lukas T.},
	Doi = {10.1007/JHEP02(2017)097},
	Eprint = {1607.06814},
	Journal = {JHEP},
	Pages = {097},
	Primaryclass = {hep-th},
	Title = {{Can Gravitational Instantons Really Constrain Axion Inflation?}},
	Volume = {02},
	Year = {2017}}

@article{Svrcek:2006yi,
	Archiveprefix = {arXiv},
	Author = {Svrcek, Peter and Witten, Edward},
	Doi = {10.1088/1126-6708/2006/06/051},
	Eprint = {hep-th/0605206},
	Journal = {JHEP},
	Pages = {051},
	Reportnumber = {SLAC-PUB-11894},
	Title = {{Axions In String Theory}},
	Volume = {06},
	Year = {2006}}

@article{Poulin:2018cxd,
	Archiveprefix = {arXiv},
	Author = {Poulin, Vivian and Smith, Tristan L. and Karwal, Tanvi and Kamionkowski, Marc},
	Doi = {10.1103/PhysRevLett.122.221301},
	Eprint = {1811.04083},
	Journal = {Phys. Rev. Lett.},
	Number = {22},
	Pages = {221301},
	Primaryclass = {astro-ph.CO},
	Title = {{Early Dark Energy Can Resolve The Hubble Tension}},
	Volume = {122},
	Year = {2019}}

@article{Poulin:2018dzj,
	Archiveprefix = {arXiv},
	Author = {Poulin, Vivian and Smith, Tristan L. and Grin, Daniel and Karwal, Tanvi and Kamionkowski, Marc},
	Doi = {10.1103/PhysRevD.98.083525},
	Eprint = {1806.10608},
	Journal = {Phys. Rev. D},
	Number = {8},
	Pages = {083525},
	Primaryclass = {astro-ph.CO},
	Title = {{Cosmological implications of ultralight axionlike fields}},
	Volume = {98},
	Year = {2018}}

@article{Harlow:2022ich,
    author = "Harlow, Daniel and Heidenreich, Ben and Reece, Matthew and Rudelius, Tom",
    title = "{Weak gravity conjecture}",
    eprint = "2201.08380",
    archivePrefix = "arXiv",
    primaryClass = "hep-th",
    reportNumber = "ACFI-T22-01",
    doi = "10.1103/RevModPhys.95.035003",
    journal = "Rev. Mod. Phys.",
    volume = "95",
    number = "3",
    pages = "035003",
    year = "2023"
}

@article{Brown:2015iha,
	Archiveprefix = {arXiv},
	Author = {Brown, Jon and Cottrell, William and Shiu, Gary and Soler, Pablo},
	Doi = {10.1007/JHEP10(2015)023},
	Eprint = {1503.04783},
	Journal = {JHEP},
	Pages = {023},
	Primaryclass = {hep-th},
	Reportnumber = {MAD-TH-15-04},
	Title = {{Fencing in the Swampland: Quantum Gravity Constraints on Large Field Inflation}},
	Volume = {10},
	Year = {2015}}

@article{Grimm:2018ohb,
	Archiveprefix = {arXiv},
	Author = {Grimm, Thomas W. and Palti, Eran and Valenzuela, Irene},
	Doi = {10.1007/JHEP08(2018)143},
	Eprint = {1802.08264},
	Journal = {JHEP},
	Pages = {143},
	Primaryclass = {hep-th},
	Title = {{Infinite Distances in Field Space and Massless Towers of States}},
	Volume = {08},
	Year = {2018}}

@article{Cheung:2014vva,
	Archiveprefix = {arXiv},
	Author = {Cheung, Clifford and Remmen, Grant N.},
	Doi = {10.1103/PhysRevLett.113.051601},
	Eprint = {1402.2287},
	Journal = {Phys. Rev. Lett.},
	Pages = {051601},
	Primaryclass = {hep-ph},
	Reportnumber = {CALT-68-2879},
	Title = {{Naturalness and the Weak Gravity Conjecture}},
	Volume = {113},
	Year = {2014}}

@article{Brown:2015lia,
	Archiveprefix = {arXiv},
	Author = {Brown, Jon and Cottrell, William and Shiu, Gary and Soler, Pablo},
	Doi = {10.1007/JHEP04(2016)017},
	Eprint = {1504.00659},
	Journal = {JHEP},
	Pages = {017},
	Primaryclass = {hep-th},
	Reportnumber = {MAD-TH-15-05},
	Title = {{On Axionic Field Ranges, Loopholes and the Weak Gravity Conjecture}},
	Volume = {04},
	Year = {2016}}

@article{Ooguri:2006in,
	Archiveprefix = {arXiv},
	Author = {Ooguri, Hirosi and Vafa, Cumrun},
	Doi = {10.1016/j.nuclphysb.2006.10.033},
	Eprint = {hep-th/0605264},
	Journal = {Nucl. Phys. B},
	Pages = {21--33},
	Reportnumber = {CALT-68-2600, HUTP-06-A017},
	Title = {{On the Geometry of the String Landscape and the Swampland}},
	Volume = {766},
	Year = {2007}}

@article{Heidenreich:2019zkl,
	Archiveprefix = {arXiv},
	Author = {Heidenreich, Ben and Reece, Matthew and Rudelius, Tom},
	Doi = {10.1007/JHEP10(2019)055},
	Eprint = {1906.02206},
	Journal = {JHEP},
	Pages = {055},
	Primaryclass = {hep-th},
	Title = {{Repulsive Forces and the Weak Gravity Conjecture}},
	Volume = {10},
	Year = {2019}}

@article{Heidenreich:2015nta,
	Archiveprefix = {arXiv},
	Author = {Heidenreich, Ben and Reece, Matthew and Rudelius, Tom},
	Doi = {10.1007/JHEP02(2016)140},
	Eprint = {1509.06374},
	Journal = {JHEP},
	Pages = {140},
	Primaryclass = {hep-th},
	Title = {{Sharpening the Weak Gravity Conjecture with Dimensional Reduction}},
	Volume = {02},
	Year = {2016}}

@article{Palti:2017elp,
	Archiveprefix = {arXiv},
	Author = {Palti, Eran},
	Doi = {10.1007/JHEP08(2017)034},
	Eprint = {1705.04328},
	Journal = {JHEP},
	Pages = {034},
	Primaryclass = {hep-th},
	Title = {{The Weak Gravity Conjecture and Scalar Fields}},
	Volume = {08},
	Year = {2017}}

@article{Montero:2015ofa,
	Archiveprefix = {arXiv},
	Author = {Montero, Miguel and Uranga, Angel M. and Valenzuela, Irene},
	Doi = {10.1007/JHEP08(2015)032},
	Eprint = {1503.03886},
	Journal = {JHEP},
	Pages = {032},
	Primaryclass = {hep-th},
	Reportnumber = {IFT-UAM-CSIC-15-028, FTUAM-15-8},
	Title = {{Transplanckian axions!?}},
	Volume = {08},
	Year = {2015}}

@article{Heidenreich:2015wga,
	Archiveprefix = {arXiv},
	Author = {Heidenreich, Ben and Reece, Matthew and Rudelius, Tom},
	Doi = {10.1007/JHEP12(2015)108},
	Eprint = {1506.03447},
	Journal = {JHEP},
	Pages = {108},
	Primaryclass = {hep-th},
	Title = {{Weak Gravity Strongly Constrains Large-Field Axion Inflation}},
	Volume = {12},
	Year = {2015}}

@article{Hebecker:2015rya,
	Archiveprefix = {arXiv},
	Author = {Hebecker, Arthur and Mangat, Patrick and Rompineve, Fabrizio and Witkowski, Lukas T.},
	Doi = {10.1016/j.physletb.2015.07.026},
	Eprint = {1503.07912},
	Journal = {Phys. Lett. B},
	Pages = {455--462},
	Primaryclass = {hep-th},
	Title = {{Winding out of the Swamp: Evading the Weak Gravity Conjecture with F-term Winding Inflation?}},
	Volume = {748},
	Year = {2015}}

@article{Bachlechner:2015qja,
	Archiveprefix = {arXiv},
	Author = {Bachlechner, Thomas C. and Long, Cody and McAllister, Liam},
	Doi = {10.1007/JHEP01(2016)091},
	Eprint = {1503.07853},
	Journal = {JHEP},
	Pages = {091},
	Primaryclass = {hep-th},
	Title = {{Planckian Axions and the Weak Gravity Conjecture}},
	Volume = {01},
	Year = {2016}}

@article{Rudelius:2021oaz,
    author = "Rudelius, Tom",
    title = "{Dimensional reduction and (Anti) de Sitter bounds}",
    eprint = "2101.11617",
    archivePrefix = "arXiv",
    primaryClass = "hep-th",
    doi = "10.1007/JHEP08(2021)041",
    journal = "JHEP",
    volume = "08",
    pages = "041",
    year = "2021"
}

@article{Andriot:2020lea,
	Archiveprefix = {arXiv},
	Author = {Andriot, David and Cribiori, Niccol\`o and Erkinger, David},
	Doi = {10.1007/JHEP07(2020)162},
	Eprint = {2004.00030},
	Journal = {JHEP},
	Pages = {162},
	Primaryclass = {hep-th},
	Title = {{The web of swampland conjectures and the TCC bound}},
	Volume = {07},
	Year = {2020}}

@article{Bedroya:2019snp,
	Archiveprefix = {arXiv},
	Author = {Bedroya, Alek and Vafa, Cumrun},
	Doi = {10.1007/JHEP09(2020)123},
	Eprint = {1909.11063},
	Journal = {JHEP},
	Pages = {123},
	Primaryclass = {hep-th},
	Title = {{Trans-Planckian Censorship and the Swampland}},
	Volume = {09},
	Year = {2020}}

@article{Gendler:2020dfp,
    author = "Gendler, Naomi and Valenzuela, Irene",
    title = "{Merging the weak gravity and distance conjectures using BPS extremal black holes}",
    eprint = "2004.10768",
    archivePrefix = "arXiv",
    primaryClass = "hep-th",
    doi = "10.1007/JHEP01(2021)176",
    journal = "JHEP",
    volume = "01",
    pages = "176",
    year = "2021"
}

@article{Lee:2019wij,
    author = "Lee, Seung-Joo and Lerche, Wolfgang and Weigand, Timo",
    title = "{Emergent strings from infinite distance limits}",
    eprint = "1910.01135",
    archivePrefix = "arXiv",
    primaryClass = "hep-th",
    reportNumber = "CERN-TH-2019-159",
    doi = "10.1007/JHEP02(2022)190",
    journal = "JHEP",
    volume = "02",
    pages = "190",
    year = "2022"
}

@article{Gonzalo:2019gjp,
    author = "Gonzalo, Eduardo and Ib{\'a}{\~n}ez, Luis E.",
    title = "{A Strong Scalar Weak Gravity Conjecture and Some Implications}",
    eprint = "1903.08878",
    archivePrefix = "arXiv",
    primaryClass = "hep-th",
    reportNumber = "IFT-UAM-CSIC-19-25",
    doi = "10.1007/JHEP08(2019)118",
    journal = "JHEP",
    volume = "08",
    pages = "118",
    year = "2019"
}

@article{Grimm:2019wtx,
    author = "Grimm, Thomas W. and Van De Heisteeg, Damian",
    title = "{Infinite Distances and the Axion Weak Gravity Conjecture}",
    eprint = "1905.00901",
    archivePrefix = "arXiv",
    primaryClass = "hep-th",
    doi = "10.1007/JHEP03(2020)020",
    journal = "JHEP",
    volume = "03",
    pages = "020",
    year = "2020"
}

@article{Hebecker:2015zss,
	Archiveprefix = {arXiv},
	Author = {Hebecker, Arthur and Rompineve, Fabrizio and Westphal, Alexander},
	Doi = {10.1007/JHEP04(2016)157},
	Eprint = {1512.03768},
	Journal = {JHEP},
	Pages = {157},
	Primaryclass = {hep-th},
	Reportnumber = {DESY-15-242},
	Slaccitation = {%%CITATION = ARXIV:1512.03768;%%},
	Title = {{Axion Monodromy and the Weak Gravity Conjecture}},
	Volume = {04},
	Year = {2016}}

@article{delaFuente:2014aca,
	Archiveprefix = {arXiv},
	Author = {de la Fuente, Anton and Saraswat, Prashant and Sundrum, Raman},
	Doi = {10.1103/PhysRevLett.114.151303},
	Eprint = {1412.3457},
	Journal = {Phys. Rev. Lett.},
	Number = {15},
	Pages = {151303},
	Primaryclass = {hep-th},
	Reportnumber = {UMD-PP-014-023},
	Slaccitation = {%%CITATION = ARXIV:1412.3457;%%},
	Title = {{Natural Inflation and Quantum Gravity}},
	Volume = {114},
	Year = {2015}}

@article{Palti:2019pca,
    author = "Palti, Eran",
    title = "{The Swampland: Introduction and Review}",
    eprint = "1903.06239",
    archivePrefix = "arXiv",
    primaryClass = "hep-th",
    reportNumber = "MPP-2019-53",
    doi = "10.1002/prop.201900037",
    journal = "Fortsch. Phys.",
    volume = "67",
    number = "6",
    pages = "1900037",
    year = "2019"
}

@article{Long:2016jvd,
    author = "Long, Cody and McAllister, Liam and Stout, John",
    title = "{Systematics of Axion Inflation in Calabi-Yau Hypersurfaces}",
    eprint = "1603.01259",
    archivePrefix = "arXiv",
    primaryClass = "hep-th",
    doi = "10.1007/JHEP02(2017)014",
    journal = "JHEP",
    volume = "02",
    pages = "014",
    year = "2017"
}

@article{Conlon:2016aea,
	Archiveprefix = {arXiv},
	Author = {Conlon, Joseph P. and Krippendorf, Sven},
	Doi = {10.1007/JHEP04(2016)085},
	Eprint = {1601.00647},
	Journal = {JHEP},
	Pages = {085},
	Primaryclass = {hep-th},
	Slaccitation = {%%CITATION = ARXIV:1601.00647;%%},
	Title = {{Axion decay constants away from the lamppost}},
	Volume = {04},
	Year = {2016}}

@article{dimopoulos:2005ac,
	Archiveprefix = {arXiv},
	Author = {Dimopoulos, S. and Kachru, S. and McGreevy, J. and Wacker, Jay G.},
	Doi = {10.1088/1475-7516/2008/08/003},
	Eprint = {hep-th/0507205},
	Journal = {JCAP},
	Pages = {003},
	Primaryclass = {hep-th},
	Reportnumber = {SLAC-PUB-11016, SU-ITP-05-08},
	Slaccitation = {%%CITATION = HEP-TH/0507205;%%},
	Title = {{N-flation}},
	Volume = {0808},
	Year = {2008}}

@article{liddle:1998jc,
	Archiveprefix = {arXiv},
	Author = {Liddle, Andrew R. and Mazumdar, Anupam and Schunck, Franz E.},
	Doi = {10.1103/PhysRevD.58.061301},
	Eprint = {astro-ph/9804177},
	Journal = {Phys. Rev.},
	Pages = {061301},
	Primaryclass = {astro-ph},
	Reportnumber = {SUSSEX-AST-98-4-3},
	Slaccitation = {%%CITATION = ASTRO-PH/9804177;%%},
	Title = {{Assisted inflation}},
	Volume = {D58},
	Year = {1998}}

@article{Rudelius:2015xta,
	Archiveprefix = {arXiv},
	Author = {Rudelius, Tom},
	Doi = {10.1088/1475-7516/2015/9/020},
	Eprint = {1503.00795},
	Journal = {JCAP},
	Pages = {020},
	Primaryclass = {hep-th},
	Slaccitation = {%%CITATION = ARXIV:1503.00795;%%},
	Title = {{Constraints on Axion Inflation from the Weak Gravity Conjecture}},
	Volume = {09},
	Year = {2015}}

@article{ArkaniHamed:2006dz,
	Archiveprefix = {arXiv},
	Author = {Arkani-Hamed, Nima and Motl, Lubos and Nicolis, Alberto and Vafa, Cumrun},
	Doi = {10.1088/1126-6708/2007/06/060},
	Eprint = {hep-th/0601001},
	Journal = {JHEP},
	Pages = {060},
	Primaryclass = {hep-th},
	Reportnumber = {HUTP-05-A0057},
	Slaccitation = {%%CITATION = HEP-TH/0601001;%%},
	Title = {{The String landscape, black holes and gravity as the weakest force}},
	Volume = {0706},
	Year = {2007}}

@article{Banks:2003sx,
	Archiveprefix = {arXiv},
	Author = {Banks, Tom and Dine, Michael and Fox, Patrick J. and Gorbatov, Elie},
	Doi = {10.1088/1475-7516/2003/06/001},
	Eprint = {hep-th/0303252},
	Journal = {JCAP},
	Pages = {001},
	Primaryclass = {hep-th},
	Reportnumber = {SCIPP-2003-03},
	Slaccitation = {%%CITATION = HEP-TH/0303252;%%},
	Title = {{On the possibility of large axion decay constants}},
	Volume = {0306},
	Year = {2003}}

@article{Rudelius:2014wla,
	Archiveprefix = {arXiv},
	Author = {Rudelius, Tom},
	Doi = {10.1088/1475-7516/2015/04/049},
	Eprint = {1409.5793},
	Journal = {JCAP},
	Number = {04},
	Pages = {049},
	Primaryclass = {hep-th},
	Slaccitation = {%%CITATION = ARXIV:1409.5793;%%},
	Title = {{On the Possibility of Large Axion Moduli Spaces}},
	Volume = {1504},
	Year = {2015}}

@article{Freese:1990rb,
	Author = {Freese, Katherine and Frieman, Joshua A. and Olinto, Angela V.},
	Doi = {10.1103/PhysRevLett.65.3233},
	Journal = {Phys.Rev.Lett.},
	Pages = {3233-3236},
	Reportnumber = {FERMILAB-PUB-90-177-A},
	Slaccitation = {%%CITATION = PRLTA,65,3233;%%},
	Title = {{Natural inflation with pseudo - Nambu-Goldstone bosons}},
	Volume = {65},
	Year = {1990}}

@article{Giddings:1987cg,
    author = "Giddings, Steven B. and Strominger, Andrew",
    title = "{Axion Induced Topology Change in Quantum Gravity and String Theory}",
    reportNumber = "HUTP-87-A067",
    doi = "10.1016/0550-3213(88)90446-4",
    journal = "Nucl. Phys. B",
    volume = "306",
    pages = "890--907",
    year = "1988"
}

@article{Bergshoeff:2004fq,
    author = "Bergshoeff, E. and Collinucci, Andres and Gran, U. and Roest, D. and Vandoren, S.",
    title = "{Non-extremal D-instantons}",
    eprint = "hep-th/0406038",
    archivePrefix = "arXiv",
    reportNumber = "KCL-MTH-04-07, SPIN-04-08, ITP-04-14, UG-04-02",
    doi = "10.1088/1126-6708/2004/10/031",
    journal = "JHEP",
    volume = "10",
    pages = "031",
    year = "2004"
}

@article{Bergshoeff:2004pg,
    author = "Bergshoeff, E. and Collinucci, Andres and Gran, U. and Roest, D. and Vandoren, S.",
    editor = "Kiritsis, E.",
    title = "{Non-extremal instantons and wormholes in string theory}",
    eprint = "hep-th/0412183",
    archivePrefix = "arXiv",
    reportNumber = "KCL-MTH-04-16, ITP-UU-04-51, SPIN-04-33, UG-04-04",
    doi = "10.1002/prop.200410227",
    journal = "Fortsch. Phys.",
    volume = "53",
    pages = "990--996",
    year = "2005"
}

@article{Arkani-Hamed:2007cpn,
    author = "Arkani-Hamed, Nima and Orgera, Jacopo and Polchinski, Joseph",
    title = "{Euclidean wormholes in string theory}",
    eprint = "0705.2768",
    archivePrefix = "arXiv",
    primaryClass = "hep-th",
    doi = "10.1088/1126-6708/2007/12/018",
    journal = "JHEP",
    volume = "12",
    pages = "018",
    year = "2007"
}

@article{Martucci:2024trp,
    author = "Martucci, Luca and Risso, Nicol{\`o} and Valenti, Alessandro and Vecchi, Luca",
    title = "{Wormholes in the axiverse, and the species scale}",
    eprint = "2404.14489",
    archivePrefix = "arXiv",
    primaryClass = "hep-th",
    doi = "10.1007/JHEP07(2024)240",
    journal = "JHEP",
    volume = "07",
    pages = "240",
    year = "2024"
}

@article{Reece:2025zva,
    author = "Reece, Matthew and Rudelius, Tom and Tudball, Christopher",
    title = "{Co-scaling and alignment of electric and magnetic towers}",
    eprint = "2505.22713",
    archivePrefix = "arXiv",
    primaryClass = "hep-th",
    doi = "10.1007/JHEP09(2025)146",
    journal = "JHEP",
    volume = "09",
    pages = "146",
    year = "2025"
}

@article{Reece:2025thc,
    author = "Reece, Matthew",
    title = "{Extra-dimensional axion expectations}",
    eprint = "2406.08543",
    archivePrefix = "arXiv",
    primaryClass = "hep-ph",
    doi = "10.1007/JHEP07(2025)130",
    journal = "JHEP",
    volume = "07",
    pages = "130",
    year = "2025"
}

@article{Heidenreich:2021yda,
    author = "Heidenreich, Ben and Reece, Matthew and Rudelius, Tom",
    title = "{The Weak Gravity Conjecture and axion strings}",
    eprint = "2108.11383",
    archivePrefix = "arXiv",
    primaryClass = "hep-th",
    reportNumber = "ACFI-T21-10",
    doi = "10.1007/JHEP11(2021)004",
    journal = "JHEP",
    volume = "11",
    pages = "004",
    year = "2021"
}

@article{Gutperle:2002km,
    author = "Gutperle, Michael and Sabra, Wafic",
    title = "{Instantons and wormholes in Minkowski and (A)dS spaces}",
    eprint = "hep-th/0206153",
    archivePrefix = "arXiv",
    reportNumber = "HUTP-02-A025, CAMS-02-03",
    doi = "10.1016/S0550-3213(02)00942-2",
    journal = "Nucl. Phys. B",
    volume = "647",
    pages = "344--356",
    year = "2002"
}

@article{Heidenreich:2016jrl,
    author = "Heidenreich, Ben and Reece, Matthew and Rudelius, Tom",
    title = "{Axion Experiments to Algebraic Geometry: Testing Quantum Gravity via the Weak Gravity Conjecture}",
    eprint = "1605.05311",
    archivePrefix = "arXiv",
    primaryClass = "hep-th",
    doi = "10.1142/S0218271816430057",
    journal = "Int. J. Mod. Phys. D",
    volume = "25",
    number = "12",
    pages = "1643005",
    year = "2016"
}

@article{Reece:2023czb,
    author = "Reece, Matthew",
    title = "{TASI Lectures: (No) Global Symmetries to Axion Physics}",
    eprint = "2304.08512",
    archivePrefix = "arXiv",
    primaryClass = "hep-ph",
    doi = "10.22323/1.439.0008",
    journal = "PoS",
    volume = "TASI2022",
    pages = "008",
    year = "2024"
}

@article{Cicoli:2021gss,
    author = "Cicoli, Michele and Guidetti, Veronica and Righi, Nicole and Westphal, Alexander",
    title = "{Fuzzy Dark Matter candidates from string theory}",
    eprint = "2110.02964",
    archivePrefix = "arXiv",
    primaryClass = "hep-th",
    reportNumber = "DESY-21-153",
    doi = "10.1007/JHEP05(2022)107",
    journal = "JHEP",
    volume = "05",
    pages = "107",
    year = "2022"
}

@article{Choi:2015zra,
    author = "Choi, Kiwoon and Chun, Eung Jin and Im, Sang Hui and Jeong, Kwang Sik",
    title = "{Diluting the inflationary axion fluctuation by a stronger QCD in the early Universe}",
    eprint = "1505.00306",
    archivePrefix = "arXiv",
    primaryClass = "hep-ph",
    reportNumber = "CTPU-15-06, KIAS-P15021",
    doi = "10.1016/j.physletb.2015.08.041",
    journal = "Phys. Lett. B",
    volume = "750",
    pages = "26--30",
    year = "2015"
}

@article{Ibe:2018ffn,
    author = "Ibe, Masahito and Yamazaki, Masahito and Yanagida, Tsutomu T.",
    title = "{Quintessence Axion Revisited in Light of Swampland Conjectures}",
    eprint = "1811.04664",
    archivePrefix = "arXiv",
    primaryClass = "hep-th",
    reportNumber = "IPMU-18-0183",
    doi = "10.1088/1361-6382/ab5197",
    journal = "Class. Quant. Grav.",
    volume = "36",
    number = "23",
    pages = "235020",
    year = "2019"
}

@article{Seo:2024zzs,
    author = "Seo, Min-Seok",
    title = "{Axion species scale and axion weak gravity conjecture-like bound}",
    eprint = "2407.16156",
    archivePrefix = "arXiv",
    primaryClass = "hep-th",
    doi = "10.1007/JHEP11(2024)082",
    journal = "JHEP",
    volume = "11",
    pages = "082",
    year = "2024"
}

@article{Shiu:2026edl,
    author = "Shiu, Gary and Tonioni, Flavio and Tran, Hung V.",
    title = "{Bounding axion dark energy}",
    eprint = "2604.09141",
    archivePrefix = "arXiv",
    primaryClass = "astro-ph.CO",
    month = "4",
    year = "2026"
}

@article{Sheridan:2024vtt,
    author = "Sheridan, Elijah and Carta, Federico and Gendler, Naomi and Jain, Mudit and Marsh, David J. E. and McAllister, Liam and Righi, Nicole and Rogers, Keir K. and Schachner, Andreas",
    title = "{Fuzzy axions and associated relics}",
    eprint = "2412.12012",
    archivePrefix = "arXiv",
    primaryClass = "hep-th",
    reportNumber = "KCL-PH-TH/2024-75, KCL-PH-TH/2024-75",
    doi = "10.1007/JHEP09(2025)016",
    journal = "JHEP",
    volume = "09",
    pages = "016",
    year = "2025"
}

@article{Stout:2020uaf,
    author = "Stout, John",
    title = "{Instanton expansions and phase transitions}",
    eprint = "2012.11605",
    archivePrefix = "arXiv",
    primaryClass = "hep-th",
    doi = "10.1007/JHEP05(2022)168",
    journal = "JHEP",
    volume = "05",
    pages = "168",
    year = "2022"
}

@article{Witten:1984dg,
    author = "Witten, Edward",
    title = "{Some Properties of O(32) Superstrings}",
    reportNumber = "Print-84-0838 (PRINCETON)",
    doi = "10.1016/0370-2693(84)90422-2",
    journal = "Phys. Lett. B",
    volume = "149",
    pages = "351--356",
    year = "1984"
}

@article{Conlon:2006tq,
    author = "Conlon, Joseph P.",
    title = "{The QCD axion and moduli stabilisation}",
    eprint = "hep-th/0602233",
    archivePrefix = "arXiv",
    reportNumber = "DAMTP-2006-17",
    doi = "10.1088/1126-6708/2006/05/078",
    journal = "JHEP",
    volume = "05",
    pages = "078",
    year = "2006"
}

@article{Andriot:2026lac,
    author = "Andriot, David",
    title = "{Dark energy from string theory: an introductory review}",
    eprint = "2603.25797",
    archivePrefix = "arXiv",
    primaryClass = "hep-th",
    month = "3",
    year = "2026"
}

@article{vandeHeisteeg:2023dlw,
    author = "van de Heisteeg, Damian and Vafa, Cumrun and Wiesner, Max and Wu, David H.",
    title = "{Species scale in diverse dimensions}",
    eprint = "2310.07213",
    archivePrefix = "arXiv",
    primaryClass = "hep-th",
    doi = "10.1007/JHEP05(2024)112",
    journal = "JHEP",
    volume = "05",
    pages = "112",
    year = "2024"
}

@article{Agmon:2022thq,
    author = "Agmon, Nathan Benjamin and Bedroya, Alek and Kang, Monica Jinwoo and Vafa, Cumrun",
    title = "{Lectures on the string landscape and the Swampland}",
    eprint = "2212.06187",
    archivePrefix = "arXiv",
    primaryClass = "hep-th",
    month = "12",
    year = "2022"
}

@article{Calderon-Infante:2022nxb,
    author = "Calder{\'o}n-Infante, Jos{\'e} and Ruiz, Ignacio and Valenzuela, Irene",
    title = "{Asymptotic accelerated expansion in string theory and the Swampland}",
    eprint = "2209.11821",
    archivePrefix = "arXiv",
    primaryClass = "hep-th",
    reportNumber = "CERN-TH-2022-153, IFT-UAM/CSIC-22-110",
    doi = "10.1007/JHEP06(2023)129",
    journal = "JHEP",
    volume = "06",
    pages = "129",
    year = "2023"
}

@article{Hamada:2019fmc,
    author = "Hamada, Yuta and Kiritsis, Elias and Nitti, Francesco and Witkowski, Lukas T.",
    title = "{Axion RG flows and the holographic dynamics of instanton densities}",
    eprint = "1905.03663",
    archivePrefix = "arXiv",
    primaryClass = "hep-th",
    reportNumber = "CCTP-2019-5, ITCP-IPP 2019/5",
    doi = "10.1088/1751-8121/ab4712",
    journal = "J. Phys. A",
    volume = "52",
    number = "45",
    pages = "454003",
    year = "2019"
}

@article{Hamada:2020phg,
    author = "Hamada, Yuta and Kiritsis, Elias and Nitti, Francesco",
    title = "{Holographic Theories at Finite {\ensuremath{\theta}}-Angle, CP-Violation, Glueball Spectra and Strong-Coupling Instabilities}",
    eprint = "2007.13535",
    archivePrefix = "arXiv",
    primaryClass = "hep-th",
    reportNumber = "CCTP-2020-9, ITCP-IPP-2020/9",
    doi = "10.1002/prop.202000111",
    journal = "Fortsch. Phys.",
    volume = "69",
    number = "2",
    pages = "2000111",
    year = "2021"
}

@article{Gaiotto:2017yup,
    author = "Gaiotto, Davide and Kapustin, Anton and Komargodski, Zohar and Seiberg, Nathan",
    title = "{Theta, Time Reversal, and Temperature}",
    eprint = "1703.00501",
    archivePrefix = "arXiv",
    primaryClass = "hep-th",
    doi = "10.1007/JHEP05(2017)091",
    journal = "JHEP",
    volume = "05",
    pages = "091",
    year = "2017"
}

@article{Dashen:1970et,
    author = "Dashen, Roger F.",
    title = "{Some features of chiral symmetry breaking}",
    doi = "10.1103/PhysRevD.3.1879",
    journal = "Phys. Rev. D",
    volume = "3",
    pages = "1879--1889",
    year = "1971"
}

@article{Witten:1980sp,
    author = "Witten, Edward",
    title = "{Large N Chiral Dynamics}",
    reportNumber = "HUTP-80/A005",
    doi = "10.1016/0003-4916(80)90325-5",
    journal = "Annals Phys.",
    volume = "128",
    pages = "363",
    year = "1980"
}

@article{Grimm:2004ua,
    author = "Grimm, Thomas W. and Louis, Jan",
    title = "{The Effective action of type IIA Calabi-Yau orientifolds}",
    eprint = "hep-th/0412277",
    archivePrefix = "arXiv",
    doi = "10.1016/j.nuclphysb.2005.04.007",
    journal = "Nucl. Phys. B",
    volume = "718",
    pages = "153--202",
    year = "2005"
}

@article{vandeHeisteeg:2023uxj,
    author = "van de Heisteeg, Damian and Vafa, Cumrun and Wiesner, Max and Wu, David H.",
    title = "{Bounds on field range for slowly varying positive potentials}",
    eprint = "2305.07701",
    archivePrefix = "arXiv",
    primaryClass = "hep-th",
    doi = "10.1007/JHEP02(2024)175",
    journal = "JHEP",
    volume = "02",
    pages = "175",
    year = "2024"
}

@article{Kaufmann:2026fli,
    author = "Kaufmann, Lukas and Monnee, Jeroen and Weigand, Timo and Wiesner, Max",
    title = "{Quantum obstructions for $N=1$ infinite distance limits -- Part I: $g_s$ obstructions}",
    eprint = "2603.12315",
    archivePrefix = "arXiv",
    primaryClass = "hep-th",
    month = "3",
    year = "2026"
}

@article{Kaufmann:2026mha,
    author = "Kaufmann, Lukas and Monnee, Jeroen and Weigand, Timo and Wiesner, Max",
    title = {{Quantum obstructions for $N=1$ infinite distance limits -- Part II: K{\"a}hler obstructions}},
    eprint = "2603.13470",
    archivePrefix = "arXiv",
    primaryClass = "hep-th",
    month = "3",
    year = "2026"
}

@article{Kaufmann:2026tsy,
    author = "Kaufmann, Lukas and Weigand, Timo and Wiesner, Max",
    title = "{On Quantum Obstructions in Type IIA Orientifolds}",
    eprint = "2604.25988",
    archivePrefix = "arXiv",
    primaryClass = "hep-th",
    month = "4",
    year = "2026"
}
\end{document}